\documentclass[a4paper,11pt,nohyper]{JHEP3}
\def\a{\alpha}

\def\c{\gamma}
\def\d{\delta}
\def\e{\epsilon}           
\def\f{\phi}               
\def\g{\gamma}

\def\i{\iota}

\def\k{\kappa}                    

\def\m{\mu}
\def\n{\nu}
\def\o{\omega}  \def\w{\omega}
\def\t{\tau}
\def\u{\upsilon}
\def\x{\xi}

\def\D{\Delta}

\def\G{\Gamma}

\def\L{\Lambda}
\def\O{\Omega}  \def\W{\Omega}
\def\P{\Pi}
\def\Q{\Theta}
\def\S{\Sigma}
\def\U{\Upsilon}
\def\X{\Xi}
\def\del{\partial}              
\let\a=\alpha  \let\g=\gamma \let\d=\delta \let\e=\epsilon
   \let\i=\iota \let\k=\kappa
 \let\m=\mu \let\n=\nu \let\x=\xi 
 \let\t=\tau  \let\f=\phi \let\c=\chi \let\y=\psi
\let\w=\omega  \let\D=\Delta \let\Q=\Theta \let\L=\Lambda
\let\X=\Xi  \let\U=\Upsilon  \let\Y=\Psi
\let\C=\Chi \let\W=\Omega   \let\G=\Gamma

\def\nn{\nonumber} \def\bd{\begin{document}} \def\ed{\end{document}}
\def\ds{\documentstyle} \let\fr=\frac \let\bl=\bigl \let\br=\bigr
\let\Br=\Bigr \let\Bl=\Bigl
\let\bm=\bibitem
\let\na=\nabla
\let\pa=\partial \let\ov=\overline
\newcommand{\be}{\begin{equation}}
\newcommand{\ee}{\end{equation}}
\def\ba{\begin{array}}
\def\ea{\end{array}}
\def\ft#1#2{{\textstyle{{\scriptstyle #1}\over {\scriptstyle #2}}}}
\def\fft#1#2{{#1 \over #2}}
\def\del{\partial}
\def\sst#1{{\scriptscriptstyle #1}}
 \def\oneone{\rlap 1\mkern4mu{\rm l}}
\def\ie{{\it i.e.\ }}
\def\via{{\it via}}
\def\semi{{\ltimes}}
\def\str{{\rm str}}
\def\Dm{{{D_{\sst{max}}}}}
\def\vac{ \left | 0 \right \rangle }
\def\kvac{ \left | k \right \rangle }

\def\sp{\; \; \;}

\def\bol{ \left | B (p^+) \right \rangle}
\def\bo1{ \left | B^0 (p^+) \right \rangle}

\def\bolt{ \left | B (p^+) \right \rangle_{\t}}

\def\boxl{ \left | B (x^-) \right \rangle}
\newcommand{\bea}{\begin{eqnarray}}
\newcommand{\eea}{\end{eqnarray}}

\def\<{ \langle }
\def\>{ \rangle }

\def\S{\Sigma}

\usepackage{amsmath,latexsym,amssymb,slashed}
\usepackage{graphicx}
\usepackage{psfrag}
\usepackage[latin1]{inputenc}
\usepackage{subfigure}

\renewcommand{\floatpagefraction}{0.6}
\renewcommand{\textfraction}{0.2}

\newcommand\ca{\mathcal{A}}
\newcommand\vp{\varphi}
\newcommand\beal{\begin{align}}
\newcommand\bbone{\ensuremath{\mathbbm{1}}}

\newcommand{\eq}[1]{\begin{equation}#1\end{equation}}
\newcommand{\spl}[1]{\begin{split}#1\end{split}}
\newcommand{\al}[1]{\begin{align}#1\end{align}}
\newcommand{\subeq}[1]{\begin{subequations}#1\end{subequations}}

\newcommand{\arXividhepth}[1]{\href{http://arxiv.org/abs/#1}arXiv:{\tt #1} [hep-th]}
\newcommand{\arXividother}[2]{\href{http://arxiv.org/abs/#1}arXiv:{\tt #1} [#2]}

\newcommand{\bg}[1]{\hat{#1}}
\newcommand{\wj}{\widetilde{J}}
\newcommand{\reo}{\mathrm{Re}~\!\omega}
\newcommand{\imo}{\mathrm{Im}~\!\omega}
\newcommand{\ads}{AdS_4}
\newcommand{\mcal}{\mathcal{M}}
\newcommand{\ccal}{\mathcal{C}}
\newcommand{\ncal}{\mathcal{N}}
\newcommand{\boxedeq}[1]{
\begin{equation}
\fbox{
\rule[0.7cm]{0pt}{0pt}
$#1$
\rule[-0.45cm]{0pt}{0pt}
}
\end{equation}
}

\def\d{\text{d}}

\def\slashchar#1{\setbox0=\hbox{$#1$}           
\dimen0=\wd0                                 
\setbox1=\hbox{/} \dimen1=\wd1               
\ifdim\dimen0>\dimen1                        
\rlap{\hbox to \dimen0{\hfil/\hfil}}      
#1                                        
\else                                        
\rlap{\hbox to \dimen1{\hfil$#1$\hfil}}   
/                                         
\fi}

\def\Re           {{\rm Re\hskip0.1em}}
\def\Im           {{\rm Im\hskip0.1em}}

\newcommand{\E}{\text{\tiny E}}

\newcommand{\tV}{{\widetilde{V}}}
\newcommand{\tH}{{\tilde{h}}}
\newcommand{\tm}{{{m}}}
\newcommand{\tmu}{{\tilde{\mu}}}
\newcommand{\trho}{{\tilde{\rho}}}
\newcommand{\tv}{{\tilde{v}}}
\newcommand{\calo}{\mbox{${\cal O}$}}
\newcommand{\cala}{\mbox{${\cal A}$}}
\newcommand{\dd}{\mathrm{d}}
\newcommand{\ra}{\rightarrow}
\newcommand{\calv}{\mbox{${\cal V}$}}
\newcommand{\calh}{\mbox{${\cal H}$}}
\newcommand{\calm}{\mbox{${\cal M}$}}
\newcommand{\abs}[1]{\left| #1 \right|}
\newcommand{\zetaa}{{\psi}}
\newcommand{\tr}{{\rm tr}\,}

\newcommand{\ky}[1]{{\color{blue}{#1}}}

\title{Entanglement entropy and differential entropy for massive flavors}

\author{Peter A. R. Jones${}^{\spadesuit}$ and Marika Taylor${}^{\diamondsuit}$  \\

\begin{itemize}
\renewcommand{\labelitemi}{${}^\spadesuit$}
\item Physics and Astronomy and STAG Research Centre, University of Southampton, \\
Highfield, Southampton, SO17 1BJ, UK.

 \renewcommand{\labelitemi}{${}^\diamondsuit$}
\item Mathematical Sciences and STAG Research Centre, University of Southampton, \\
Highfield, Southampton, SO17 1BJ, UK.

  \end{itemize}

\bigskip
 E-mail:
 \email{p.jones@soton.ac.uk; m.m.taylor@soton.ac.uk}}

\abstract{In this paper we compute the holographic entanglement entropy for massive flavors in the D3-D7 system, for arbitrary mass and various entangling region geometries. We show that the universal terms in the entanglement entropy exactly match those computed in the dual theory using conformal perturbation theory. We derive holographically the universal terms in the entanglement entropy for a CFT perturbed by a relevant operator, up to second order in the coupling; our results are valid for any entangling region geometry. We present a new method for computing the entanglement entropy of any top-down brane probe system using Kaluza-Klein holography and illustrate our results with massive flavors at finite density. Finally we discuss the differential entropy for brane probe systems, emphasising that the differential entropy captures only the effective lower-dimensional Einstein metric rather than the ten-dimensional geometry.}

\begin{document}

\newcommand{\td}{\tilde}
 \newcommand{\bc}{\begin{center}}
 \newcommand{\ec}{\end{center}}
 \newcommand{\bfr}{\begin{flushright}}
 \newcommand{\efr}{\end{flushright}}
 \newcommand{\bfl}{\begin{flushleft}}
 \newcommand{\efl}{\end{flushleft}}
 \newcommand{\bt}{\begin{tabular}}
 \newcommand{\et}{\end{tabular}}

\section{Introduction}

In recent years there has been considerable interest in entanglement entropy and its holographic realisation, following the proposal of
\cite{Ryu:2006bv} that entanglement entropy can be computed from the area of a bulk minimal surface
cohomologous to a boundary entangling region. This proposal was proved for spherical entangling regions in \cite{Casini:2011kv} and
arguments supporting the Ryu-Takayanagi prescription based on generalised entropy were given in \cite{Lewkowycz:2013nqa}.  Entanglement entropy has been computed in a wide range of holographic systems, see the review \cite{Takayanagi:2012kg}. 

The focus of this paper is on the computation of holographic entanglement entropy in top-down brane probe systems, which are widely used in phenomenological applications of holography. Entanglement entropy is a new computable for such systems and, following the pioneering works of  \cite{Klebanov:2007ws,Huijse:2011ef}, can act as an order parameter for confinement and other phase transitions. 

A top-down brane probe system is expressed in terms of a ten-dimensional supergravity background and a brane embedding into this background. The Ryu-Takayanagi prescription is however based on extremal surfaces in the 
reduced Einstein $(d+1)$-dimensional metric, where $d$ is the dimension of the dual field theory. One of the main results of this paper is a systematic method to compute the holographic entanglement entropy for any top-down brane probe system, using the method of Kaluza-Klein holography \cite{Skenderis:2006uy} to extract the lower-dimensional Einstein metric. This method reproduces earlier results of \cite{Jensen:2013lxa,Chang:2013mca,Karch:2014ufa,Estes:2014hka} but allows entanglement entropy to be computed for any brane system with arbitrary worldvolume fluxes. (Earlier results for massless flavors at finite density can be found in \cite{Chang:2014oia}.)
We illustrate our methodology using the example of the D3-D7 system at finite mass and density. We compute the holographic entanglement entropy for massive flavors, with arbitrary mass and various entangling region geometries, and use our new methodology to address the case of finite density. 

Brane systems provide a new testing ground for the dependence of entanglement entropy on the field theory and on the shape of entangling region, topics of considerable current interest, see for example \cite{Klebanov:2012yf,Nishioka:2014kpa,Lee:2014zaa,Allais:2014ata,Mezei:2014zla,Ben-Ami:2015zsa}. In particular, one can explore the structure of universal logarithmic terms; these are well-understood for conformal field theories (see e.g. \cite{Schwimmer:2008yh,Solodukhin:2008dh}) and recent papers have explored the behaviour of entanglement entropy under relevant perturbations using conformal perturbation theory 
\cite{Rosenhaus:2014nha,Rosenhaus:2014ula,Rosenhaus:2014zza,He:2014lfa,Casini:2014yca,Park:2015dia}. It was shown in 
\cite{Rosenhaus:2014zza} that for a CFT deformed by a relevant operator
\be
I \rightarrow I + \lambda \int d^{d} x {\cal O}
\ee
there is a new logarithmic divergence in the entanglement entropy of the half space
\be
\delta S = {\cal N} \lambda^2 \frac{(d-2)}{4 (d-1)} \frac{\pi^{\frac{d+2}{2}}}{\Gamma ( \frac{d+2}{2})} {\cal A} \log \left ( \frac{\e_{UV}}{\e_{IR}} \right ) \label{prov2}
\ee
when $\Delta = (d+2)/2$ with ${\cal A}$ the area of the dividing surface and ${\cal N}$ the normalisation of the two point function of the operator ${\cal O}$. Here $\e_{UV}$ and $\e_{IR}$ correspond to UV and IR cutoffs respectively. 

In section \ref{six} we prove \eqref{prov2} by analysing the volume divergences of the holographic entanglement entropy and show that (as postulated in \cite{Rosenhaus:2014zza}) such a divergence occurs for an entangling surface with arbitrary geometry. 
We also show explicitly that \eqref{prov2} agrees with the logarithmic terms in the entanglement entropy for the D3-D7 system at finite mass, using the results of \cite{Karch:2005ms} to determine the holographic two point function normalisation. 
As well as matching the universal terms in the entanglement entropy, we explain the origin of finite terms in the entanglement entropy for massive flavor systems, in terms of the effective IR description of the system in terms of a CFT deformed by irrelevant operators. 

\bigskip

There is a growing literature connecting quantum entanglement with the global structure of the bulk spacetime, see in particular  \cite{Maldacena:2013xja,Swingle:2014uza}. In \cite{Bianchi:2012ev} a relation between the area of generic (non-minimal) surfaces and entanglement was proposed and this idea was sharpened with the introduction of differential entropy \cite{Balasubramanian:2013rqa,Balasubramanian:2013lsa,Myers:2014jia,Balasubramanian:2014sra,Czech:2014wka,Headrick:2014eia}. We verify that the differential entropy in the D3-D7 system indeed computes the area of a hole in the reduced Einstein metric; the agreement is somewhat subtle since the depth of the hole is itself corrected by the presence of the probe branes. 

In section \ref{eight} we discuss the implications of the fact that the entanglement and differential entropy are related to the reduced Einstein metric, rather than the ten-dimensional metric: even if entanglement allows us to reconstruct the reduced Einstein metric completely, this information does not suffice to reconstruct the ten-dimensional geometry. Moreover, the causal structure in ten dimensions  only agrees with that of the reduced Einstein metric in special cases (e.g. product metrics); the global structure is qualitatively different between five and ten dimensions even for well-understood examples such as the Coulomb branch of ${\cal N} = 4$ SYM. Reconstruction of the full ten-dimensional geometry would therefore seem to require a generalized notion of entanglement in the dual field theory.

\bigskip

The plan of the paper is as follows. In section \ref{two} we briefly review the relevant features of the D3-D7 system. 
In sections \ref{three} and \ref{four} we compute the entanglement entropy for slab, half space and spherical entangling regions for the massive D3-D7 system. In section \ref{kk} we present a general method to compute the entanglement entropy in any brane probe system using Kaluza-Klein holography and illustrate our method with the D3-D7 system at finite mass and density. We discuss the field theory interpretation of our results in section \ref{six} and give a holographic proof of \eqref{prov2} for generic entangling regions. In section \ref{seven} we show that the differential entropy computes the area of a hole in the Einstein metric and we discuss the meaning of entanglement and differential entropy for top-down solutions in section \ref{eight}, illustrating our discussions with Coulomb branch geometries. We conclude in section \ref{nine} and various technical results relevant to section \ref{kk} are contained in appendix \ref{ten}.

\section{Massive flavors} \label{two}
 
In this paper we will explore entanglement entropy for massive brane systems, focussing for the most part on the specific example of the D3-D7 brane system. Consider $N_c$ D3-branes and $N_f \ll N_c$ parallel coincident D7-branes. As discussed in the early days of the AdS/CFT correspondence \cite{Fayyazuddin:1998fb,Aharony:1998xz} the decoupling limit gives rise to ${\cal N} = 4$ SYM coupled to $N_f$ massless flavors; the resulting field theory is an ${\cal N} = 2$ SCFT.  Taking the background $AdS_5 \times S^5$ metric to be of unit radius:
\be
ds^2 = \frac{1}{z^2}\left (dz^2 + dx^{\mu}dx^{\mu} \right ) + d \theta^2 + \sin^2 \theta d \Omega_3^2 + \cos^2 \theta d \phi^2
\ee
the embedding of a probe D7-brane corresponding to a massless flavor is described by $\phi$ constant and $\theta = \pi/2$, i.e. the probe D7-brane wraps
$AdS_5 \times S^3$. 

Suppose one separates the D7-branes from the stack of D3-branes; the resulting open strings are massive and the field theory in the decoupling limit corresponds to 
 ${\cal N} = 4$ SYM coupled to $N_f$ massive flavors \cite{Karch:2002sh} (we discuss the massive deformation of the ${\cal N} =2$ SCFT further in section \ref{six}). The corresponding D7-brane embedding in $AdS_5 \times S^5$ is described by $\phi$ being constant and the angle $\theta$ depending on the radial coordinate $z$ as 
\be
\sin^2 \theta = (1 - m^2 z^2),
\ee
where $m$ corresponds to the flavor mass, or equivalently the separation of the D7 and D3 branes. Note that the probe brane wraps the equator of the $S^5$ as $z \rightarrow 0$ and smoothly caps off at a finite value of $z = 1/m$, controlled by the flavor mass. 

The D3-D7 brane system can be used to model mesons holographically. Considerable work has been done on generalizations of the probe brane embeddings to finite temperature and finite density, see the review \cite{Erdmenger:2007cm} and on the backreaction of the flavor branes onto the geometry \cite{Bigazzi:2009bk,Bigazzi:2011it,Bigazzi:2011db}. In particular, note that interesting meson melting phase transitions are observed at finite temperature and density, see for example \cite{Mateos:2006nu,Kobayashi:2006sb}. Backreacting massive flavors is non-trivial even at zero temperature and density, since the flavors break the global symmetry to $SO(4)$ and the resulting ten-dimensional metric is therefore of cohomogeneity three. Smearing the branes over the compact space reduces the cohomogeneity of the metric but this is obscure from the field theory perspective as it corresponds to an averaging over different field theories. 

In this work we will calculate the entanglement entropy and the differential entropy for the massive flavor system at zero temperature and zero density, and match our results with field theory results based on conformal perturbation theory. We will also present a method to compute the entanglement entropy for any probe brane system (with or without worldvolume gauge fields) and illustrate this method with the case of massive flavors at finite density. The method is equally applicable at finite temperature, although at finite temperature the entanglement entropy will include both thermal and quantum contribution; matching with field theory results is considerably harder as few results for finite temperature exist. It would however be interesting to explore the finite temperature results in the context of melting phase transitions.

\section{Entanglement entropy for slabs} \label{three}

\begin{figure}
\begin{center}
\setlength{\unitlength}{0.50mm}
\includegraphics*[width=0.4\linewidth]{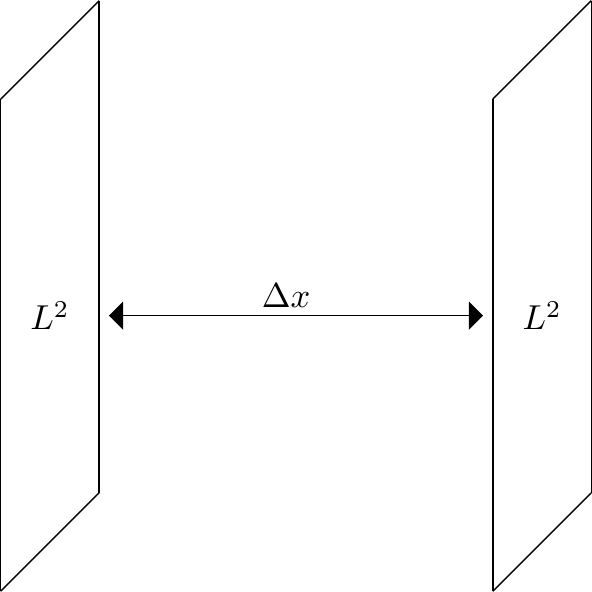}
\caption{Illustration of a slab boundary region: $\Delta x =l$ is the width of the region, and $L^2$ is the regularized area of its boundary faces.}
\label{fig:slab}
\end{center}
\end{figure} 

In this section we compute the entanglement entropy for a slab on the boundary (see Figure \ref{fig:slab}), working to leading order in the ratio of the number of flavors to colors, ${N}_f/ N_c$.

Let us begin by reviewing the computation of entanglement entropy for a slab in $AdS_{5}$. 
We define a slab region on the boundary of width $\Delta x =l$ by $x \in [0,l]$, and take as an embedding ansatz $z=z(x)$ on a $t=0$ hypersurface. Defining the regularised lengths of the other spatial directions as $L$, it is then easy to show that the Ryu-Takayanagi entanglement entropy functional \cite{Ryu:2006bv} for the embedding surface is:
\begin{equation}
S =\frac{L^2}{4G_N} \int_{0}^{l}dx\frac{\sqrt{1+z'^2}}{z^3},
\end{equation}
where $G_{N}$ is the Newton constant. Since we chose the $AdS_5$ to have unit radius, the Newton constant is dimensionless and can be related to the number of colors $N_c$ as
\be
\frac{1}{8 \pi G_N} = \frac{N_c^2}{4 \pi^2}.
\ee 
The Lagrangian is independent of $x$ explicitly and hence the associated Hamiltonian is a constant of motion. Rearranging the expression for this constant of motion one easily finds:
\begin{equation}
z'=\frac{\sqrt{\tilde{z}^6-z^6}}{z^3}
\end{equation}
where $\tilde{z}$ is clearly the turning point of the solution since $z'(\tilde{z})=0$. The entanglement entropy is then obtained by substituting this solution into the entropy functional, resulting in
\be
S = \frac{L^2 \tilde{z}^3}{2G_{N}} \int^{\tilde{z}}_{\epsilon} \frac{ dz}{z^3 \sqrt{\tilde{z}^6 - z^6}}.
\ee
Here we have included a factor of two, from the two halves of the entangling surface, i.e. $0 < x < l/2$ and $l/2< x < l$. 
It is useful to define the dimensionless parameter $s\equiv z/\tilde{z}$ so that $z \in [\epsilon,\tilde{z}] \rightarrow s \in [a,1]$ where $a \equiv \epsilon/\tilde{z}$ is also dimensionless. We obtain, for example:
\begin{equation}
\frac{dx}{dz}=\frac{s^3}{\sqrt{1-s^6}}
\end{equation}
Note that the entanglement entropy is thereby manifestly dimensionless
\bea
S &=& \frac{L^2}{2 \tilde{z}^2 G_N}  \int^{1}_{a} \frac{ ds}{s^3 \sqrt{1 - s^6}}  \label{ads5-s} \\
&=& \frac{L^2 }{2 \tilde{z}^2 G_N} \left ( \frac{\sqrt{\pi } \Gamma(-\frac{1}{3})}{6 \Gamma( \frac{1}{6})} + 
\frac{1}{2 a^2} {}_2F_1\left(-1/3,1/2,2/3,a^6\right) \right ); \nn \\ 
&=& \frac{ L^2}{2G_N} \left ( \frac{1}{2 \e^2} + \frac{\sqrt{\pi} \Gamma( -\frac{1}{3})}{6 \Gamma ( \frac{1}{6}) \tilde{z}^2} \right ), \nn
\eea
where in the latter equation we retain only terms which are finite or divergent as the cutoff $\e \rightarrow 0$. 

\begin{figure}
\begin{center}
\setlength{\unitlength}{0.50mm}
\includegraphics*[width=0.7\linewidth]{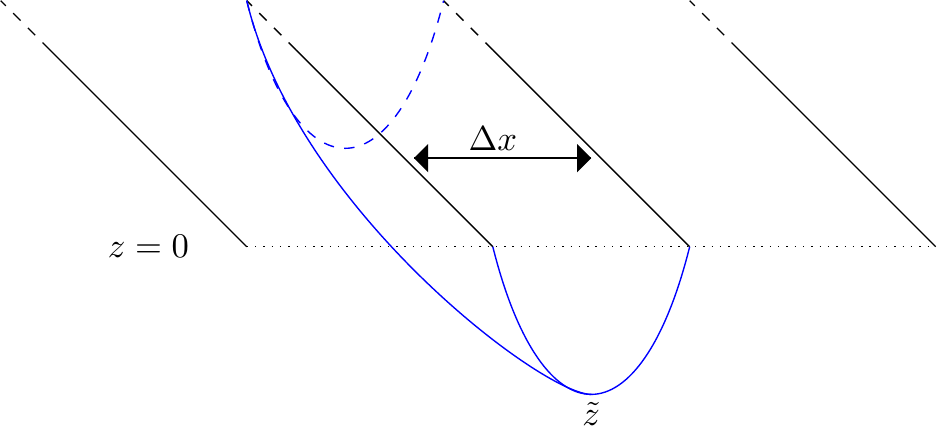}
\caption{The minimal surface for a slab boundary region - the boundary is at $z=0$ and $z=\tilde{z}$ is the turning point of the surface.}
\label{fig:slabsurface}
\end{center}
\end{figure} 

It is simple to find the induced metric of the entangling surface and its associated stress tensor. The induced metric is given by $\gamma^{\min}_{ab}=\partial_a X^{\mu}\partial_b X^{\nu}g_{\mu \nu}$ where $a,b=(s,y,w)$ run over the surface indices and $\mu,\nu$ run over all AdS$_5$ indices. The induced metric is therefore
\begin{equation}
\gamma^{\min}_{ab}= \left(\frac{1}{s^2(1-s^6)},\frac{1}{s^2 \tilde{z}^2},\frac{1}{s^2 \tilde{z}^2}\right)
\end{equation}
We note for further use that
\begin{equation}
\sqrt{\gamma^{\min}}=\frac{1}{\tilde{z}^2s^3\sqrt{1-s^6}}.
\end{equation}

Differentiating the action functional, the stress tensor for the surface is given by:
\begin{equation}
T^{\mu \nu}_{min}\equiv \frac{2}{\sqrt{\gamma^{\min}}}\frac{\delta \sqrt{\gamma^{\min}}}{\delta g_{\mu \nu}} 
\end{equation}
which evaluates to:
\begin{equation}
T^{\mu \nu}_{min}=\gamma^{\min ab}\partial_a X^{\mu}\partial_b X^{\nu}
\end{equation}
after using the chain rule. It is then a simple matter to calculate these components, resulting in:
\begin{equation}
T^{\mu \nu}_{min}=\left( s^2(1-s^6)\tilde{z}^2,0,s^8\tilde{z}^2,s^2\tilde{z}^2,s^2\tilde{z}^2 \right)
\end{equation}
which we will make use of below.  Note also that the relation between the width of the slab, $l$, and the turning point of the minimal surface is
\be
l = 2 \int_{0}^{\tilde{z}} \frac{z^3 dz}{(\tilde{z}^6 - z^6)^{\frac{1}{2}}}   = 2 \tilde{z} \int_{0}^1 \frac{s^3 ds}{(1 - s^6)^{\frac{1}{2}}} = \frac{2 \sqrt{\pi} \Gamma ( \frac{2}{3} )}{\Gamma ( \frac{1}{6}) } \tilde{z}.  \label{rel1}
\ee

\subsection{Flavor contribution}

Now let us compute the change in the entanglement entropy caused by the presence of ${N}_f$ flavors branes, with ${ N}_f \ll {N}_c$. A priori, to compute this change one would expect that one needs to compute the backreaction of the branes to linear order in ${N}_{f}/{N}_c$ and then extract from the backreacted geometry the change in the five-dimensional Einstein metric and hence the change in the area of the minimal surface. As mentioned earlier, it is hard to compute the backreacted ten-dimensional geometry because of the high cohomogeneity of the problem; smeared solutions are known and entanglement entropy was computed for these smeared solutions in \cite{Kontoudi:2013rla}. 

It is important to note however that entanglement entropy is defined in terms of the five-dimensional Einstein metric, not the ten-dimensional (Einstein-frame) metric. One does not in general obtain the correct answer for the entanglement entropy by computing the area of a minimal surface in the ten-dimensional metric, see section \ref{kk}. Note that \cite{Kontoudi:2013rla} used the ten-dimensional metric rather than the five-dimensional Einstein metric. 

Computing the full ten-dimensional backreaction without smearing and extracting the effective five-dimensional Einstein metric is intractable for general probe brane systems. Several methods have therefore been developed to extract the entanglement entropy from the probe brane embedding, see \cite{Jensen:2013lxa,Chang:2013mca,Karch:2014ufa,Estes:2014hka}. The methods of \cite{Jensen:2013lxa,Karch:2014ufa,Estes:2014hka} are particularly applicable to spherical entangling regions, for which the CHM map \cite{Casini:2011kv} may be exploited. In this section our discussion will follow that of \cite{Chang:2013mca}, which is applicable to all entangling region geometries. 

On general grounds the change in the entanglement entropy for any perturbation in the five-dimensional Einstein metric is 
\begin{equation}
\delta S=\frac{1}{4G_N}\int d^3 x \sqrt{\gamma^{\min}} \frac{1}{2}T^{\mu \nu}_{\min} h^E_{\mu \nu} \label{pert-ee}
\end{equation}
where $h^E_{\mu \nu}$ is the perturbation in the five-dimensional Einstein metric, $T^{\mu \nu}_{\min}$ is the energy momentum tensor of the minimal (entangling) surface in the background and the integral is over the original entangling surface. Therefore one can compute the entanglement entropy provided one can extract the change in the five-dimensional Einstein metric. For general brane embeddings the computation of the perturbation in the five-dimensional Einstein metric is subtle; in section \ref{kk} we present a method to compute the Einstein metric for all types of brane embeddings. 

It was observed in \cite{Chang:2013mca} that the perturbation in the five-dimensional Einstein metric is straightforward to compute whenever the brane embedding has an induced worldvolume metric which is diagonal (a product of a non-compact part and a compact part which is embedded in the sphere part of the background geometry) and the non-compact part of the metric has no dependence on the sphere coordinates. 
In such a case the linearised backreaction on the metric for probe branes can be computed \cite{Karch:2014ufa} as:
\begin{equation}
h^E_{\m\n} = \frac{1}{z^2} {\rm diag} \left(f(z),-h(z),h(z),h(z),h(z)\right) \label{metric-p}
\end{equation}
where the metric perturbation is sourced by the effective brane energy momentum tensor $T^{{\rm eff}}_{\m\n}$ i.e.
\be
G_{\m\n} (h^E) = 8 \pi G_{N} T^{ {\rm eff}}_{\m\n}. \label{p-ein}
\ee
This effective stress energy tensor is obtained by reducing the brane action over the three-sphere:
\be
I = - T_7 \int_{AdS_5} d^5 \sigma \int_{S^3} d^3 \sigma \sqrt{-\gamma} = - T_7 \int_{AdS_5}  d^5 \sigma  (2 \pi^2)(1 - m^2 z^2)^{\frac{3}{2}} \sqrt{-\gamma_{(s)}},
\ee
where $\gamma_{\alpha \beta}$ is the worldvolume metric for the brane\footnote{Note that we denote the worldvolume metric for the brane as $\gamma$ and the induced metric on the entangling surface as $\gamma^{\min}$.}. The worldvolume metric is diagonal  for the given embedding and therefore the determinant factorises, allowing the integral over the three-sphere to be evaluated. The resulting effective action then depends only on the non-compact part of the worldvolume metric $\gamma_{(s) \m \n}$, but note that the effective tension of this brane is $z$ dependent. 
Varying this effective action with respect to the non-compact part of the background metric results in the effective energy momentum tensor
\be
(T^{{\rm eff}})^{\m \n} = 2 \pi^2 T_7 (1 - m^2 z^2) \gamma_{s}^{\m\n}. \label{t-source}
\ee
Note that this method for computing the effective source term for the five-dimensional Einstein metric relies on the fact that the worldvolume brane metric is a direct product of non-compact and compact parts. The method is also not applicable for brane embeddings in which worldvolume gauge fields are non-zero or worldvolume fields source other supergravity fields as well as the metric. In section \ref{kk} we will discuss a more generally applicable method for computing the entanglement entropy contributions from probe branes which does not rely on a diagonal worldvolume metric. 

Substituting \eqref{t-source}  into \eqref{p-ein} gives the following equation:
\begin{equation} 
f(z) + z h'(z) \equiv \td{f}(z) = \frac{t_0}{12}(1-m^2 z^2)^2 \label{g-inv}
\end{equation} 
Here $t_0$ is the backreaction parameter, proportional to the number of flavors ${\cal N}_f$:
\be
t_0 = 16 \pi G_N T_o; \qquad T_o = 2 \pi^2 T_7 \label{tens}
\ee
where $T_7$ is the tension of a D7-brane. Only the gauge invariant combination $\td{f}(z)$ is determined by the Einstein equations. However, continuity of the metric and of the extrinsic curvature at $z=1/m$ requires that $h(z)$ satisfies $h(1/m)=h'(1/m)=0$.

\begin{figure}
\begin{center}
\setlength{\unitlength}{0.50mm}
\includegraphics*[width=0.7\linewidth]{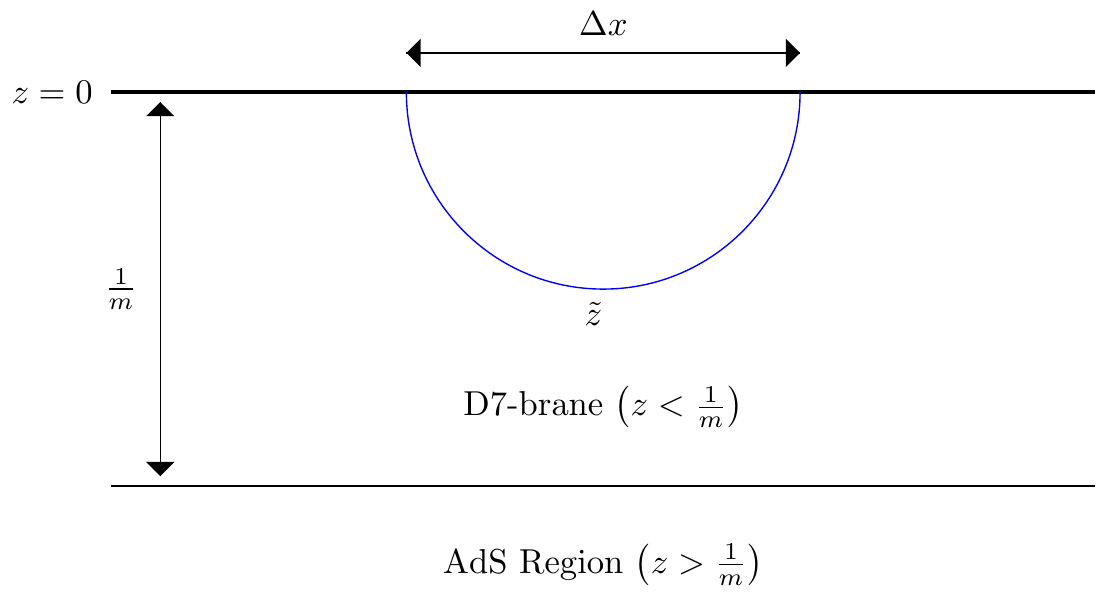}
\caption{The relationship between the original minimal surface and the D7-probe embedding - when $\tilde{z} < 1/m$ the entire minimal surface lies within the probe brane embedding.}
\label{fig:minsurface}
\end{center}
\end{figure} 

Substituting the metric perturbation and the minimal surface stress energy tensor into \eqref{pert-ee} thus gives
\begin{equation}
\delta S=\frac{1}{4G_N}\int ds dw dy \frac{1}{2 \tilde{z}^2s^3 \sqrt{1-s^6}}\left(f(\tilde{z}s)(1-s^6)+h(\tilde{z}s)(s^6+2)\right)
\end{equation}
where this integral is over the original entangling surface with coordinates $(s,w,y)$. Defining $\alpha \equiv L^2/(4G_N\tilde{z}^2)$ for convenience (where we have computed the trivial $y,w$-integrals to give the factor $L^2$ and taken into account the factor of two arising from the two halves of the entangling surface), and unpacking $f(\tilde{z}s)$ we find:
\begin{equation}
\delta S=\alpha \int ds \left(\frac{\sqrt{1-s^6}}{s^3}\tilde{f}(\tilde{z}s)-\frac{\sqrt{1-s^6}}{s^2}\tilde{z}h'(\tilde{z}s)+\frac{(s^6+2)}{s^3\sqrt{1-s^6}}h(\tilde{z}s)\right)
\end{equation}
Integrating the second term by parts, the bulk contribution cancels the third term and one is left with a boundary contribution:
\begin{equation}
\delta S=\alpha \left(\int_a^b ds \frac{\sqrt{1-s^6}}{s^3} \tilde{f}(\tilde{z}s)- \left[h(\tilde{z}s)\frac{\sqrt{1-s^6}}{s^2} \right]^{s=b}_{s=a}\right) \label{final-ee}
\end{equation}
where $a=\epsilon/\tilde{z}$ and $b=1,1/\mu$ for $\mu<1, \mu>1$ respectively. Here $\mu \equiv m \tilde{z}$ is the dimensionsless mass parameter. This latter distinction occurs because, although the integral runs over the original entangling surface which has $z \in [\epsilon,\tilde{z}]$, the integral in fact only receives a non-zero contribution when $h^E_{\mu \nu}\neq 0$, i.e. for $z < 1/m$. When $\tilde{z}<1/m$ (i.e. $\mu<1$), the entire entangling surface lies within the brane embedding, and the upper limit is thus $z=\tilde{z}$ or $s=1$. When $\tilde{z}>1/m$ (i.e. $\mu>1$) however, the upper limit will depend on the mass and be given by $z=1/m$ i.e. $s=1/\mu$ (see Figure \ref{fig:minsurface}). 

For both cases, the boundary term at $s=b$ actually vanishes. The expression within square brackets trivially vanishes at $s=1$  for the case $\mu < 1$, and it vanishes for $\mu \ge 1$ using the continuity condition $h(1/m)=0$. The expression for both cases is thus given by:
\begin{equation}
\delta S=\alpha \left(\int_a^b ds \frac{\sqrt{1-s^6}}{s^3} \tilde{f}
(\tilde{z}s)+ h(\epsilon)\frac{\sqrt{1-a^6}}{a^2} \right)
\end{equation}
with $b$ depending on the case as mentioned above. 
Expanding for $\tilde{f}$ this becomes
\begin{equation}
\delta S= \frac{t_0\alpha}{12} \left(\int_a^b ds \frac{\sqrt{1-s^6}}{s^3}(1-\mu^2 s^2)^2 + h(\epsilon)\frac{\sqrt{1-a^6}}{a^2} \right)
\end{equation}
Note that the entanglement entropy depends explicitly on the gauge fixing for the metric perturbation.  One choice of scheme would be to set $f(z) = 0$, corresponding to Fefferman-Graham coordinates while a second natural choice of scheme is to fix $h(z)$ such that the cutoff is unchanged to linear order and one then obtains the relation:
\begin{equation}
h(\epsilon)= \frac{t_0}{12}\left(1-\frac{2}{3}m^2 \epsilon^2+\frac{1}{5}m^4 \epsilon^4 \right)+\mathcal{O}(t_0^2)
\end{equation}
For this gauge choice one obtains 
\bea
\delta S &=& \frac{t_0 L^2}{48 G_N} \left( \frac{1}{\tilde{z}^2} \int_a^b ds \frac{\sqrt{1-s^6}}{s^3}(1-\mu^2 s^2)^2+ \left(\frac{1}{\epsilon^2}-\frac{2 m^2}{3} +\mathcal{O}(\epsilon^2)\right) \right); \nn \\
&=& \frac{t_0 L^2}{48 G_N} \left( \frac{1}{\tilde{z}^2} \int_a^b ds \frac{\sqrt{1-s^6}}{s^3}(1-\mu^2 s^2)^2 \right) + \delta S_{\rm gauge} (m, \epsilon), \label{smallm}
\eea
where we note that the gauge dependent contribution $\delta S_{\rm gauge}$ is independent of the turning point $\tilde{z}$, since $h(0)$ is finite. We will discuss this point further below. In what follows we will retain the gauge dependence explicitly, rather than fixing a gauge, and show that this gauge dependence drops out of universal terms.

In computing the integral we first specialise to the case of small mass so $b=1$. Performing the integral directly over the range $s \in [a,1]$ and expanding the answer in $a$ gives the following up to $\mathcal{O}(a)$:
\begin{equation}
\int_a^1 ds \frac{\sqrt{1-s^6}}{s^3}(1-\mu^2 s^2)^2 = \frac{1}{2a^2}+\frac{2}{3}\mu^2+\frac{\sqrt{\pi}}{12}\frac{\Gamma(-1/3)}{\Gamma(7/6)}+\mu^4\frac{\sqrt{\pi}}{12}\frac{\Gamma(1/3)}{\Gamma(11/6)}-\frac{2}{3}\mu^2 \textrm{log}2+2\mu^2 \textrm{log}a
\end{equation}
The result thus becomes:
\bea
\delta S &=& \frac{t_0L^2}{48 G_N} \left(\frac{1}{2 \epsilon^2} + \frac{2}{3} m^2 +\frac{\sqrt{\pi}} {12 \tilde{z}^2} \frac{\Gamma(-1/3)}{\Gamma(7/6)}+  m^4 \tilde{z}^2 \frac{\sqrt{\pi}}{12} \frac{\Gamma(1/3)}{\Gamma(11/6)} + \frac{2}{3} m^2 \textrm{log} (\epsilon^3/2\tilde{z}^3) \right) \nn  \\ && 
+ \delta S_{\rm gauge} (m,\epsilon). \label{smallmstrip}
\eea
We next consider the case of large mass so $b=1/\mu$. The  result is given in terms of generalised hypergeometric functions:
\bea
\int_a^{1/\mu} ds \frac{\sqrt{1-s^6}}{s^3}(1-\mu^2 s^2)^2 &=& \frac{1}{2a^2} +\frac{1}{6\mu^4} {}_3F_2\left(\{1/2,1,1\},\{2,2\},1/\mu^6\right) \label{midm} \\
&& -\frac{\mu^2}{2}{}_2F_1\left(-1/2,-1/3,2/3,1/\mu^6\right) \nn \\
&& +\frac{\mu^2}{2}{}_2F_1\left(-1/2,1/3,4/3,1/\mu^6\right)+2\mu^2 \textrm{log} (\mu a) +\mathcal{O}(a^2) \nn
\eea
Expanding for large mass one then obtains
\begin{equation}
\delta S = \frac{t_0L^2}{48 G_N } \Bigg(\frac{1}{2 \epsilon^2} + 2 m^2 \textrm{log} (m \epsilon)  - \frac{1}{48 m^4 \tilde{z}^6} +\mathcal{O}\left(\frac{\epsilon^2}{\tilde{z}^2}\right) + \mathcal{O}\left(\frac{1}{m^{10} \tilde{z}^{12}}\right) \Bigg) + \delta S_{\rm gauge}(m, \epsilon).  \label{largem}
\end{equation}
Note that the power and log-divergent terms agree for $\mu \leq 1$ and $\mu \geq 1$. 

\bigskip

We can immediately obtain the change in the entanglement entropy for the half space from the $\tilde{z} \rightarrow \infty$ limit of the above expression. In this case the entangling surface extends throughout the bulk and has no turning point. The contribution to the entanglement entropy from the brane is then
\be
\delta S = \frac{t_0L^2}{48 G_N } \Bigg(\frac{1}{4 \epsilon^2} + m^2 \textrm{log} (m \epsilon)  \Bigg ). \label{infinitel}
\ee
Note that the divergent terms differ from \eqref{largem} by an overall factor of two, since the entangling surface in the field theory no longer has two disconnected parts. We will discuss the field theory computation of \eqref{infinitel} in section 6. 

\subsection{Changes in turning point and entanglement surface}

\begin{figure}
\begin{center}
\setlength{\unitlength}{0.80mm}
\includegraphics*[width=0.7\linewidth]{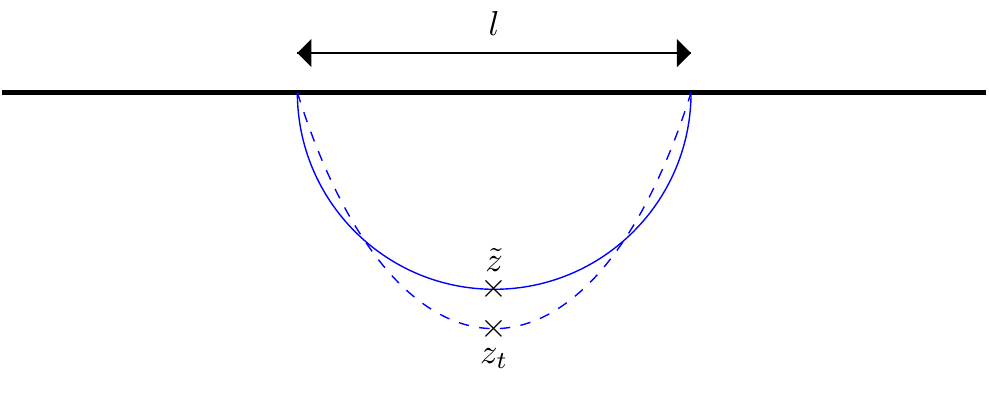}
\caption{Illustration of the change in the turning point of the minimal surface - $\tilde{z}$ is the turning point of the original minimal surface (i.e. the one that is actually used to compute the flavor contribution to the entanglement entropy (\ref{pert-ee})), whereas $z_t$ is the turning point of the minimal surface in the backreacted geometry.}
\label{fig:turning}
\end{center}
\end{figure} 

The perturbed entangling surface has a turning point for which the relation between the turning point and the width of the slab $\Delta x = l$ is changed relative to (3.12). The equation for the perturbed entangling surface is obtained analogously to (3.3) and given by
\be
(z')^2 = \left ( \frac{\tilde{z}^6}{z^6} - 1 \right ) + h(z) \left ( \frac{4 \tilde{z}^6}{z^6} - 1 \right ) - f(z) \left ( \frac{\tilde{z}^6}{z^6} - 1 \right ). \label{sur-cor}
\ee
for some constant $\tilde{z}$, and where $f(z)$ and $h(z)$ are the metric perturbations discussed previously. The width of the slab is then given by 
\be
\frac{l}{2} =  \int_{0}^{\tilde{z}} \frac{z^3 dz}{ ( \tilde{z}^6 - z^6)^{\frac{1}{2}}} + \frac{1}{2} \int_{0}^{\frac{1}{m}} z^3 dz 
\frac{(f + z h')}{ (\tilde{z}^6 - z^6)^{\frac{1}{2}}} - \frac{1}{2} \left [ \frac{z^4 h}{ (\tilde{z}^6 - z^6)^{\frac{1}{2}}} \right ]^{\frac{1}{m}}_{0},  \label{int-sur}
\ee
where the upper limits of integration are explained as follows: for $\mu \ge 1$, the surface ends in the region in which the perturbations $f(z)$ and $h(z)$ vanish; from  \eqref{sur-cor} the turning point therefore remains at $z = \tilde{z}$, and we continue to use the definition $\mu \equiv m \tilde{z}$.  

The boundary term in \eqref{int-sur} vanishes since $h(1/m) = 0$ and $h(0)$ is finite and therefore the relation between the turning point and the slab width depends only on the gauge invariant combination of metric perturbations. Substituting this combination using \eqref{g-inv} one hence obtains
\be
\frac{l}{2} = \tilde{z} \left ( \frac{\sqrt{\pi} \Gamma(\frac{2}{3})}{\Gamma( \frac{1}{6})}  + \frac{t_0}{24} \int_{0}^{\mu^{-1}}
ds s^3 \frac{(1 - \mu^2 s^2)^2}{(1 - s^6)^{\frac{1}{2}}} \right ),  \label{c1largem1}
\ee
where note that $\mu = m \tilde{z}$ depends on $\tilde{z}$ implicitly. The integral can be computed resulting in 
\bea
\int_{0}^{\mu^{-1}}
ds s^3 \frac{(1 - \mu^2 s^2)^2}{(1 - s^6)^{\frac{1}{2}}}  &=& \frac{1}{24 \mu^4} \left ( 16 \mu^3 ( -\mu^3 + \sqrt{\mu^6-1}) 
+ 6 \; {}_2F_1\left(\frac{1}{2},\frac{2}{3},\frac{5}{3},\frac{1}{\mu^6}\right) \right .\label{c1largem2} \\ 
&& \left . \qquad \qquad + 3 \; {}_2F_{1}\left( \frac{1}{2},\frac{4}{3},\frac{7}{3},\frac{1}{\mu^6}\right) \right ) \nn 
\eea
and expanding for $\mu \gg 1$ one finds
\be
\frac{l}{2} = \tilde{z} \left ( \frac{\sqrt{\pi} \Gamma(\frac{2}{3})}{\Gamma( \frac{1}{6})} + \frac{t_0}{576 m^4 \tilde{z}^4} \right ).  \label{rel2}
\ee

\bigskip

For the small mass case the situation is more complicated, since from equation \eqref{sur-cor} we find that the turning point of the surface is itself changed (see Figure \ref{fig:turning}). Let the perturbed turning point be
\be
z_{t} = \tilde{z} + t_{0} \delta \tilde{z}.
\ee
The latter is computed by setting $z' = 0$ in \eqref{sur-cor}, resulting in 
\be
t_0 \delta \tilde{z} = \frac{1}{2} \tilde{z} h(\tilde{z}).  \label{turning-point}
\ee
Note that the shift in the turning point depends on the metric perturbation $h(z)$ explicitly, rather than the gauge independent combination. 

The relation between $l$ and $z_t$ is now calculated using the relation \eqref{sur-cor} which can be rewritten as
\be
z^3 z' = \left (  ( 1+ h(\tilde{z}) )(z_t^6 -z^6) - f(z) (z_t^6 - z^6) + H(z) (4 \tilde{z}^6 - z^6) \right )^{\frac{1}{2}}, \label{sur-rel2}
\ee
where we define
\be
h(z) = H(z) + h(\tilde{z}),
\ee
with $H(\tilde{z})$ by construction being zero. In relation \eqref{sur-rel2} we implicitly work to first order in $t_0$, which in particular implies that $\tilde{z}$ can be replaced by $z_t$ in terms multiplying $f(z)$ and $H(z)$, which are already of order $t_0$. Therefore 
\be
\frac{l}{2} =  \frac{1}{(1 + h(\tilde{z}))^{\frac{1}{2}}} \int_{0}^{z_t} \frac{z^3 dz}{ ( z_t^6 - z^6)^{\frac{1}{2}}} + \frac{1}{2} \int_{0}^{z_t} z^3 dz 
\frac{(f + z H')}{ (z_t^6 - z^6)^{\frac{1}{2}}} - \frac{1}{2} \left [ \frac{z^4 H}{ (\tilde{z}^6 - z^6)^{\frac{1}{2}}} \right ]^{z_t}_{0}. 
\ee
The boundary term vanishes at $z=0$ and the contribution at $z=z_t$ is zero since $H(z_t) = 0$ to order $t_0$. Since $h'(z) = H'(z)$ the combination appearing in the second integral is the gauge invariant combination as before and therefore
\be
\frac{l}{2} = z_t ( 1 - \frac{1}{2} h(\tilde{z}) ) \int_{0}^{1} \frac{s^3 ds}{(1 - s^6)^{\frac{1}{2}}} + \frac{z_t}{24} t_0 \int_{0}^1 ds \frac{s^3 (1 - \mu^2  s^2)^2}{(1 -s^6)^{\frac{1}{2}}}.  
\ee
where now $\mu \equiv m z_t$. Computing the integrals we obtain
\bea
l &=& z_t ( 1 - \frac{1}{2} h(\tilde{z}) ) \frac{2 \sqrt{\pi} \Gamma(\frac{2}{3})}{\Gamma(\frac{1}{6})} \label{lzrel2}  \\ 
&& \qquad + \frac{t_0 z_t}{12} \left  ( \frac{\sqrt{\pi} \Gamma(\frac{2}{3})}{\Gamma(\frac{1}{6})}  - \frac{2\mu^2}{3}  + \frac{\mu^4 \sqrt{\pi} \Gamma(\frac{4}{3})}{6 \Gamma( \frac{11}{6})} \right ).  \nn
\eea
Note that $h(\tilde{z}) = h(z_t)$ at this order. 

Even though we only needed the original entangling surface to compute the entanglement entropy above, these changes to the turning point are important to keep track of when comparing the differential entropy to the gravitational entropy of the corresponding hole in the bulk, as we will discuss in section 7.

\subsection{Finite contributions}

To understand the infra-red behaviour of the entanglement entropy it is often useful to isolate the finite contributions. 

A dimensionless, cut-off independent quantity was defined in \cite{Hertzberg:2010uv,Hertzberg:2012mn} by differentiating with respect to the mass as
\be
S_{m} = m^4 \frac{\partial^2 S}{\partial (m^2)^2}; \label{sm}
\ee
note that this expression is valid for four-dimensional quantum field theories, with different expressions being proposed in lower dimensions. Implicitly, $\epsilon$ and $\tilde{z}$ (or equivalently $l$) are held fixed. A priori, since the gauge dependent terms depend on the mass it is not obvious that this quantity will be independent of the gauge. However, 
on general grounds the gauge dependent terms must make the form 
\be
\delta S_{\rm gauge} = a_{-2} \frac{L^2}{\epsilon^2} + a_{0} m^2 L^2 + {\mathcal O} (\epsilon^2)
\ee
with $a_{-2}$ and $a_{0}$ dimensionless coefficients. This form follows from the fact that 
the entanglement entropy is extensive, i.e. it is proportional to the area of each slab $L^2$, and the underlying theory is conformal. This implies that the only finite terms in the scheme dependent part of the entanglement entropy must be proportional to $m^2 L^2$, since $m$ is the only other cutoff independent scale in the problem. Since neither $a_{-2}$ nor $a_0$ contribute to \eqref{sm}, the quantity computed by \eqref{sm} is indeed independent of the gauge. 

Computing this quantity one finds that for $\mu \le 1$
\be
\delta S_{m} = \frac{t_0 L^2}{48 G_N } \Bigg( \frac{\sqrt{\pi} \Gamma( \frac{1}{3})}{6 \Gamma( \frac{11}{6})} m^4 \tilde{z}^2    \Bigg ) \label{margue}
\ee
while for $\mu \ge 1$
\begin{equation}
\delta S_m = \frac{t_0 L^2}{32G_N} \frac{\mu^2}{3 \tilde{z}^2}{}_2 F_1[-1/2,1/3,4/3,1/\mu^6]
\end{equation}
which can be expanded for $\mu \gg 1$
\be
\delta S_{m} = \frac{t_0 L^2}{48 G_N } \Bigg( m^2 - \frac{1}{8 m^4 \tilde{z}^6}  + \mathcal{O}\left (\frac{1}{m^{10} \tilde{z}^{12}} \right ) \Bigg ) \label{marguel}
\ee

For slab geometries an alternative method of defining a cut-off independent quantity is (see for example \cite{Casini:2004bw,Casini:2005rm,Casini:2005zv,Casini:2009sr}\footnote{$S_l$ is always positive and decreasing in two dimensions and plays the role of a c-function.})
\be
S_{l} = l \frac{\partial S}{\partial l}. \label{sl}
\ee
This quantity is manifestly independent of the coordinate choice $h(z)$, since $\delta S_{\rm gauge}$ is a local quantity, which is hence independent of the (non-local) slab width $l$, as we saw explicitly below (\ref{smallm}). For $\mu \le 1$ this quantity evaluates to
\be
\delta S_{l} =  \frac{t_0 L^2}{48 G_N } \Bigg( - \frac{\sqrt{\pi}}{6 \tilde{z}^2} \frac{\Gamma(-\frac{1}{3})}{\Gamma(\frac{7}{6})} + \frac{\sqrt{\pi} \Gamma( \frac{1}{3})}{6 \Gamma( \frac{11}{6})} m^4 \tilde{z}^2    \Bigg ) \label{largues}
\ee
while for $\mu \gg 1$ one obtains
\be
\delta S_{l} =  \frac{t_0 L^2}{48 G_N } \left (  \frac{1}{8 m^4 \tilde{z}^6}  + \mathcal{O}\left (\frac{1}{m^{10} \tilde{z}^{12}} \right ) \right ). \label{largue}
\ee
The limit $\mu \gg 1$ probes the IR of the theory: for fixed $m$ this corresponds to taking an entangling surface which extends deep into the bulk. Therefore a finite quantity should in this limit decouple from UV physics.
Comparing \eqref{marguel} and \eqref{largue}, the first quantity does not fulfil this criterion (as the term of order $m^2$ derives from the logarithmic divergence) whereas the latter quantity does. We will hence use \eqref{largue} in section 6 when discussing the IR physics. 

\subsection{Phase transitions}

The flavor contributions to the entanglement entropy \eqref{smallmstrip} and \eqref{midm} match at $m \tilde{z} = 1$, i.e. when the turning point of the entangling surface is at the location where the metric perturbation vanishes. This matching was guaranteed by the continuity of the metric perturbation and its derivative at $z=1/m$. 
There are however discontinuities in the derivatives of the flavor contribution at $m \tilde{z}=1$, induced by the discontinuities of higher derivatives of the metric perturbation at $z= 1/m$. 
In particular, there is a discontinuity in the fourth derivative of the entanglement entropy with respect to the slab width (at fixed mass):
\be
\left (\frac{\partial^4 S}{\partial l^4} \right )_{m \tilde{z} =1}
\ee
From the field theory perspective it is more natural to fix the mass (i.e. the theory) and vary the slab width (i.e. the entangling region). However, if one instead looks at the variation of the entanglement entropy with respect to the mass at fixed slab width, the fourth derivative is also discontinuous:
\be
\left (\frac{\partial^4 S}{\partial m^4} \right )_{m \tilde{z} =1}
\ee
Correspondingly the finite quantities $\delta S_m$ and $\delta S_l$ have discontinuities in their second and third derivatives, respectively. 

The discontinuity arises from the discontinuity in second derivatives of the metric at $z = 1/m$. The fourth derivative of the entanglement entropy contains the terms:
\be
\left (\frac{\partial^4 S}{\partial l^4} \right )_{\tilde{z} =1/m} \sim \frac{1}{8 G_N} \left (\frac{\partial \tilde{z}}{\partial l} \right )^4_{\tilde{z} =1/m}
\frac{\partial^3}{\partial \tilde{z}^3} \left ( \sqrt{\gamma} T^{\mu \nu}_{\rm min} h_{\mu \nu}^E(\tilde{z}) \right )_{\tilde{z} =1/m}
\ee
Note that the discontinuity does not arise at lower order in derivatives since the volume element of the entangling surface vanishes at the turning point. 

\bigskip

At first sight one might try to assign a physical interpretation to the discontinuity of the entanglement entropy, i.e. a phase transition. However, the discontinuity is inherited from the discontinuity in the metric derivatives and correspondingly in the curvature. This discontinuity is likely to be an artefact of the probe approximation: in a fully back-reacted solution for the D3 and D7 branes there should be no source terms in the energy momentum tensor and hence no discontinuities in the curvature of the metric. In other words, one would expect from gauge/gravity duality that the backreacted solution should solve the type IIB equations with no sources. The metric and curvature should hence be continuous and the metric for the back-reacted solution should be smoothened around $z=1/m$, over a radial coordinate range $\Delta z \ll 1/m$. 

It is interesting to note that at any finite density the brane probe extends throughout the bulk and therefore there is no longer any discontinuity at finite $z$: as we show in section \ref{kk} the backreacted solution is indeed smoothened around $z = 1/m$, over a small but finite radial coordinate range.

\section{Entanglement entropy for spherical regions} \label{four}

In this section we compute the entanglement entropy for the case of a spherical entangling surface, extending the small mass results of \cite{Karch:2014ufa} to generic mass. The methods of \cite{Jensen:2013lxa,Karch:2014ufa,Estes:2014hka} are in principle applicable to spherical entangling regions but in practice the CHM map \cite{Casini:2011kv} becomes intractable for finite mass, as the probe brane embedding in the hyperbolic black hole is extremely complicated.  Therefore again our discussion will follow closely the method of \cite{Chang:2013mca}. 

Writing the boundary metric in spherical coordinates we have:
\begin{equation}
ds^2=\frac{1}{z^2}(-dt^2+dr^2+r^2 d\Omega_2^2+ dz^2)
\end{equation}
for the $AdS_5$ metric. We define a ball on the boundary by $r\leq R$ and take as an embedding ansatz $z=z(r)$ at $t=0$. The functional for the entangling surface is then:
\begin{equation}
S = \frac{\pi}{G_N}  \int_0^R dr \frac{\sqrt{1+z'^2}}{z^3}r^2,
\end{equation}
where we have done the trivial integral over the two-sphere.
It is easy to show that the resulting equations of motion are solved by the hemisphere $r^2+z^2=R^2$, and the desired extremal surface is thus a hemisphere in $r$ and $z$ of radius $R$ that wraps the 2-sphere $S^2$ - this surface will be parametrised by $\{s,\Omega_2\}$ where $s$ is defined by $z=R s$ and $r=R \sqrt{1-s^2}$ and $\Omega_2=(\theta,\phi)$.

We can now compute the induced metric on this extremal surface using $\gamma_{ab}^{min}=\partial_a X^{\mu}\partial_b X^{\nu}g_{\mu \nu}$ where now $a,b=(s,\theta,\phi)$. One finds:
\begin{equation}
\gamma_{ab}^{min}= \left(\frac{1}{s^2(1-s^2)},\frac{(1-s^2)}{s^2}g_{S^2}\right)
\end{equation}
where we note for further use that:
\begin{equation}
\sqrt{\gamma^{min}} =\frac{\sqrt{1-s^2}}{s^3}\sqrt{g_{S^2}}
\end{equation}
We can then compute the stress tensor of the surface and one finds (computing only the diagonal components since this quantity will be contracted with $h^E_{\mu\nu}$ which is diagonal):
\begin{equation}
T^{\mu \nu}_{min}=\left( s^2 R^2(1-s^2),0,R^2 s^4,\frac{s^2}{1-s^2},\frac{s^2}{1-s^2}\textrm{cosec}^2(\theta) \right)
\end{equation}

We are now in a position to compute the entanglement entropy, but we must first write the metric backreaction in the coordinate system $(z,t,r,\theta,\phi)$:
\begin{equation}
h^{E}_{\mu \nu} = \frac{1}{z^2}\textrm{diag}\left(f(z),h(z),h(z),h(z)g_{S^2} \right)
\end{equation}
The resulting entanglement entropy becomes
\begin{equation}
\delta S = \frac{\pi}{2G_N}\int_{a}^b ds \frac{\sqrt{1-s^2}}{s^3}\left( (s^2+2)h( R s)-(s^2-1)f( R s)    \right)
\end{equation}
where $a \equiv \epsilon/R$ and $b=1,1/\mu$ for $\mu<1, \mu>1$ respectively as for the slab, where now $\mu \equiv m R$. Expanding out $f(R s)$  and again using partial integration on the $h'(Rs)$ term as in the slab case one obtains:
\begin{equation}
\delta S = \frac{ \pi}{2 G_N}\int_{a}^b ds \frac{(1-s^2)^{3/2}}{s^3}\tilde{f}(Rs) - \frac{ \pi}{2G_N}\left[ h(R s) \frac{(1-s^2)^{3/2}}{s^2}\right]_a^b  \label{integrand}
\end{equation}
again reducing the contribution of $h(z)$ to a boundary term. The term at $s=b$ vanishes for both possible values of $b$ for the same reasons as before, leading to:
\begin{equation}
\delta S = \frac{t_0\pi}{24G_N}\int_{a}^b ds \frac{(1-s^2)^{3/2}}{s^3} \left(1-(\mu s)^2\right)^2  + \frac{ \pi}{2G_N} h(\epsilon) \frac{1}{a^2}\left(1-a^2\right)^{3/2} 
\end{equation}
Since $h(\epsilon)$ depends upon the gauge choice, we can rewrite this expression as in the previous section as
\be
\delta S = \frac{t_0\pi}{24G_N}\int_{a}^b ds \frac{(1-s^2)^{3/2}}{s^3} \left(1-(\mu s)^2\right)^2 + \delta S_{\rm gauge}(\e, R,m).
\ee
The gauge dependent contribution depends in this case on all three parameters: the cutoff $\epsilon$, the mass $m$ and the radius of the spherical region $R$ (note that in the previous expression the mass dependence is contained implicitly in the metric function $h(z)$). Note the difference relative to the case of the slab: since the dual theory is local, the gauge dependent terms for the slab cannot depend on the slab width. The radius of the sphere however relates to the intrinsic curvature of the entangling region, which is a local quantity and therefore can appear in the gauge dependent terms. In particular, since $h(0)$ is finite, the non-vanishing terms in $\delta S_{\textrm{gauge}}$ will be either quadratic in $R$ or independent of $R$ in any scheme. 

Let us now consider the small and large mass cases separately. 
For $\mu \le 1$ the contribution from the $s=b$ limit to the integral  vanishes where $b=1$. One therefore obtains
\be
\delta S = \frac{t_0 \pi}{8 G_N}\Bigg(\frac{R^2}{6 \epsilon^2}+\frac{4\mu^2+3}{6}\textrm{log}\frac{\epsilon}{2 R}+\frac{1}{4}+\frac{8\mu^2}{9}+\frac{\mu^4}{15}\Bigg) + \delta S_{\rm gauge}(\e,R,m). 
\ee
Using the same regularisation scheme as before to fix $h(\epsilon)$ one obtains as in \cite{Karch:2014ufa}:
\begin{equation}
\delta S = \frac{t_0 \pi}{8 G_N}\Bigg(\frac{R^2}{2 \epsilon^2}+\frac{4\mu^2+3}{6}\textrm{log}\frac{\epsilon}{2R}-\frac{1}{4}+\frac{2\mu^2}{3}+\frac{\mu^4}{15}\Bigg),
\end{equation}
where
\be
\delta S_{\rm gauge} ( \e,R,m) = \frac{\pi t_0}{24 G_N} \left ( \frac{R^2}{\e^2} - \frac{2}{3} \mu^2 - \frac{3}{2} + \cdots \right ) \label{gaugesphere}
\ee

For $\mu \ge 1$ the extra contribution from the $s=b$ limit of integration is given by:
\begin{equation}
\frac{t_0 \pi}{24 G_N}\left( -\sqrt{1-\frac{1}{\mu^2}}\left( \frac{8}{15}+\frac{83 \mu^2}{30} + \frac{\mu^4}{5} \right) + \frac{1}{2}(3+4\mu^2)\textrm{log}\left(\mu+ \mu\sqrt{1-\frac{1}{\mu^2}} \right) \right)
\end{equation}
This vanishes at $\mu=1$ as we would expect by continuity. Thus the total contribution to the entanglement entropy for $\mu \ge 1$ is given by:
\bea
\delta S &=& \frac{t_0 \pi }{8 G_N}\Bigg(\frac{R^2}{6 \e^2}+\frac{4\mu^2+3}{6}\textrm{log}\frac{\epsilon}{2 R} + \frac{1}{4}+\frac{8\mu^2}{9}+\frac{\mu^4}{15} \\
&& \qquad - \frac{1}{3}\sqrt{1-\frac{1}{\mu^2}}\left( \frac{8}{15}+\frac{83 \mu^2}{30} + \frac{\mu^4}{5} \right) + \frac{1}{6}(3+4\mu^2)\textrm{log}\left(\mu+ \mu\sqrt{1-\frac{1}{\mu^2}} \right) \Bigg) \nn \\
&& \qquad + \delta S_{\rm gauge}(\e,R,m). \nn
\eea
For $\mu\gg 1$ this expression asymptotes to 
\be
\delta S = \frac{t_0 \pi}{8 G_N} \left ( \frac{R^2}{6 \epsilon^2} + \left (\frac{2 \mu^2}{3} + \frac{1}{2} \right )  \textrm{log} (m \e) + \frac{3}{8} - \frac{1}{48 \mu^2} +  \mathcal{O} \left( \frac{1}{\mu^4}\right) \right ) + \delta S_{\rm gauge}(\epsilon,R,m). 
\ee
As for the slab, the expressions for the entanglement entropy match at $\mu = 1$, i.e. when the turning point of the entangling surface reaches $z = 1/m$. Derivatives of the entanglement entropy with respect to $R$ at fixed $m$ or with respect to $m$ at fixed $R$ become discontinuous at $\mu =1$ because of the metric discontinuity. For spherical entangling surfaces the discontinuity arises at fifth order i.e. 
\be
\left ( \frac{\partial^5 S}{\partial R^5} \right )_{m=1/R}
\ee
is discontinuous. The discontinuity again arises from the discontinuity in second derivatives of the metric at $z = 1/m$, and is hence expected to be absent in a fully back-reacted solution without sources. The fifth derivative of the entanglement entropy contains the terms:
\be
\left (\frac{\partial^4 S}{\partial l^4} \right )_{\tilde{z} =1/m} \sim \frac{1}{8 G_N} \frac{\partial^4}{\partial \tilde{z}^4} \left ( \sqrt{\gamma} T^{\mu \nu}_{\rm min} h_{\mu \nu}^E(\tilde{z}) \right )_{\tilde{z} =1/m}
\ee
Note that the discontinuity does not arise at lower order in derivatives since the terms contracted with the metric perturbation 
are zero when the turning point lies at $\tilde{z} = 1/m$, and their first derivative is also zero; see the form of the integrand in 
 \eqref{integrand}.

\subsection{Finite contributions}

For a spherical region one can define a finite quantity by differentiating with respect to the mass, \eqref{sm}. For $\mu \le 1$ this gives
\be
\delta S_{m} =  \frac{\pi t_0}{60 G_{N}} \mu^4. \label{sph1}
\ee
For $\mu \gg 1$ one obtains
\be
\delta S_{m} = - \frac{\pi t_0}{8 G_N} \left ( \frac{\mu^2}{3} - \frac{1}{4} - \frac{1}{24 \mu^2} + \mathcal{O} \left( \frac{1}{\mu^4}\right) \right ). \label{sph2}
\ee
The quantity \eqref{sl} is not finite for a spherical region and it is proposed to use instead \cite{Liu:2012eea,Liu:2013una}
\be
S_{LM} = R \frac{\partial}{\partial R} \left ( R \frac{\partial}{\partial R} - 2 \right ) S.
\ee
Note that this quantity vanishes for all terms which are independent of $R$ or quadratic in $R$, which in particular guarantees that the gauge dependent terms drop out of $S_{LM}$. 
For $\mu \le 1$ one obtains
\be
\delta S_{LM} =  \frac{\pi t_0}{8 G_N} \left ( 1 - \frac{4}{3} \mu^2 + \frac{8}{15} \mu^4  \right ). \label{sph3}
\ee
For $\mu \gg 1$ one obtains
\be
\delta S_{LM} =  \frac{\pi t_0}{4 8 G_N \mu^2} + \cdots, \label{sph4}
\ee
with all terms with higher order powers in $\mu$ cancelling. 

As for the slab
the limit $\mu \gg 1$ probes the IR of the theory: for fixed $m$ this corresponds to taking an entangling surface which extends deep into the bulk. Therefore a finite quantity should in this limit decouple from UV physics.
Comparing \eqref{sph2} and \eqref{sph4}, the first quantity again does not fulfil this criterion (as the term of order $m^2$ derives from the logarithmic divergence) whereas the latter quantity does. We will hence use \eqref{sph4} in section 6 when discussing the IR physics.


\section{Entanglement entropy from Kaluza-Klein holography} \label{kk}

In this section we describe a new method for computing the entanglement entropy of probe brane systems using Kaluza-Klein holography \cite{Skenderis:2006uy}. This method is applicable to any probe brane system, i.e. for any shape entangling region with any worldvolume gauge fields, and can also be used for other systems such as Coulomb branch geometries. 

The holographic entanglement entropy for any static asymptotically anti-de Sitter geometry is given by the Ryu-Takayanagi functional in terms of the area of a minimal surface in the Einstein frame metric. Probe brane systems are however described by top down constructions. In other words, we first specify a ten-dimensional supergravity solution for which a holographic interpretation is known, the usual examples being geometries which are asymptotic to the products of anti-de Sitter and Sasaki-Einstein manifolds. The probe system is then specified by the brane embedding into the ten-dimensional background and the worldvolume fields on the brane. The backreaction onto the ten-dimensional supergravity solution is  computed by viewing the D-brane action as sourcing the supergravity fields,  with the sources being localised on the brane embedding. Computation of the backreaction therefore involves solving all of the ten-dimensional supergravity equations. 

Even after computing the backreacted ten-dimensional supergravity solution, one cannot immediately compute the entanglement entropy, because the latter requires the five-dimensional Einstein metric. For any supergravity solution which can be viewed as a perturbation of anti-de Sitter cross a Sasaki-Einstein manifold the method of Kaluza-Klein holography can however be used to extract the five-dimensional Einstein metric  \cite{Skenderis:2006uy}. This method implies that, if one only wishes to compute the entanglement entropy, it is not actually  necessary to compute all of the backreaction of the brane onto the ten-dimersional supergravity fields: one only needs to know the backreaction for those field components which  contribute to the five-dimensional Einstein metric. 

In the rest of the section we will describe the computation of the entanglement entropy using the Kaluza-Klein holography approach for massive D7-branes in an $AdS_5 \times S^5$ background. At the end of the section we will discuss further applications and generalisations of this method. 

\subsection{Kaluza-Klein holography}

The backreaction of the D7-branes onto $AdS_5 \times S^5$ results in a supergravity background which can be expressed as a perturbation of $AdS_5 \times S^5$. Thus the metric can be expressed as 
\bea
ds^2 &=& (g^{o}_{MN} + h_{MN}) dx^M dx^N; \\
&=& 
\frac{1}{z^2} (dz^2 + dx^{\mu} dx_{\mu})  + (d \theta^2 + \sin^2 \theta d \Omega_3^2 + \cos^2 \theta d \phi^2) +  h_{MN}(x^m,\theta_a) dx^M dx^N, \nn
\eea
where we denote ten-dimensional indices as $x^M$; $\theta^a$ collectively denote the five sphere coordinates and $x^m$ denote the five-dimensional coordinates, i.e. $(z,x^{\mu})$. 
Thus $g^o_{MN}$ is the background $AdS_5 \times S^5$ metric and $h_{MN}$ is the metric perturbation. The other type IIB supergravity fields are the dilaton $\phi$, the NS-NS three form field strength $H_{MNP}$  and the RR field strengths $F_{M}$, $F_{MNP}$ and $F_{MNPQR}$. Only the self-dual five-form field strength has a background profile:
\bea
F_{MNPQR} &=& F^{o}_{MNPQR} + f_{MNPQR}; \\
F^o &=& \frac{1}{z^5} dz \wedge dt \wedge dw \wedge dx \wedge dy +  \sin^3 \theta \cos \theta d \theta \wedge d \Omega_3 \wedge d \phi, \nn
\eea
with $f_{MNPQR}$ being the perturbation of the five form field strength. Our normalisation conventions are that the Einstein equations for type IIB supergravity are given by
\be
R_{MN} = \frac{1}{6} F_{MPQRS} F_{N}^{\; PQRS} + \cdots
\ee
The Einstein equations are quadratic in the dilaton gradients and form field strengths. Therefore, working to linearized order in the perturbations, the Einstein equation decouples from the perturbations of $\phi$, $H_{MNP}$, $F_{M}$ and $F_{MNP}$ since the latter do not have profiles in the $AdS_5 \times S^5$ background. Similarly the only contributions to the five-form equation of motion at linearised order are from the metric perturbations and the five-form field strength perturbations. 

The fluctuations can be expanded in $S^5$ harmonics \cite{Kim:1985ez}:
\bea \label{fluct_h}
h_{m n}(x,y) &=& \sum h^{I_1}_{m n}(x) Y^{I_1}(y) \nonumber \\
h_{m a} (x,y)&=& 
\sum (B^{I_5}_{(v)m}(x) Y_a^{I_5}(y) 
+ B^{I_1}_{(s) m}(x) D_a Y^{I_1}(y)) \nonumber \\
h_{(ab)}(x,y) 
&=& \sum (\phi_{(t)}^{I_{14}}(x) Y_{(ab)}^{I_{14}}(y) 
+ \phi^{I_5}_{(v)}(x) D_{(a} Y^{I_5}_{b)}(y)
+ \phi^{I_1}_{(s)}(x) D_{(a} D_{b)} Y^{I_1}(y) )\nonumber \\
h_{a}^a(x,y) &=& \sum \pi^{I_1}(x) Y^{I_1}(y) 
\eea
and 
\bea
f_{m n r s t}(x,y) &=& \sum
5 D_{[m} b^{I_1}_{n r s t]}(x) Y^{I_1}(y) \nonumber \\
f_{a m n r s}(x,y) &=& \sum (b^{I_1}_{m n r s}(x) D_a Y^{I_1}(y) 
+ 4 D_{[m} b^{I_5}_{n r s]}(x) Y_a^{I_5}(y)) \nonumber \\
f_{ab m n r}(x,y) &=& \sum
(3 D_{[m} b^{I_{10}}_{n r]}(x) Y_{[ab]}^{I_{10}}(y) 
-2 b^{I_5}_{m n r}(x) D_{[a} Y_{b]}^{I_5}(y)) \nonumber \\
f_{a b c m n}(x,y) &=& \sum 
(2 D_{[m} b_{n]}^{I_5}(x) \e_{abc}{}^{de} D_d Y_e^{I_5}(y)
+ 3 b_{m n}^{I_{10}}(x) D_{[a} Y_{bc]}^{I_{10}}(y)) \nonumber \\
f_{abcd m}(x,y)  &=& \sum 
(D_m b_{(s)}^{I_1}(x) \e_{abcd}{}^e D_e Y^{I_1}(y)
+ (\L^{I_5}-4) b_m^{I_5}(x) \e_{abcd}{}^e Y_e^{I_5}(y)) \nn \\
f_{a b c d e}(x,y) &=& \sum b_{(s)}^{I_1}(x) \L^{I_1} \e_{abcde} Y^{I_1}(y) 
\eea
Numerical constants in these expressions are inserted so 
as to match with the conventions of \cite{Kim:1985ez}.
Parentheses denote a symmetric traceless combination
(i.e. $A_{(ab)} = 1/2 (A_{ab}+A_{ba}) -1/5 g_{ab} A_a^a$).
$Y^{I_1}, Y_a^{I_5}, Y_{(ab)}^{I_{14}}$ and $Y_{[ab]}^{I_{10}}$ denote scalar, 
vector and tensor harmonics whilst $\L^{I_1}$ and $\L^{I_5}$ are the 
eigenvalues of the scalar and vector harmonics under (minus) the d'Alembertian.
The subscripts $t$, $v$ and $s$ denote whether the field is associated with
tensor, vector or scalar harmonics respectively, whilst the
superscript of the harmonic label $I_n$ derives from the number of
components $n$ of the harmonic.

Not all fluctuations are independent - some are diffeomorphic to each other or to the background. This issue
can be dealt with by imposing a gauge; for example, the de Donder-Lorentz gauge fixing condition is
\be
D^{a} h_{(ab)} = D^{a} h_{a m} = 0 
\ee
which sets to zero the coefficients $B_{(s) m}^{I_1}, \f_{(v)}^{I_5}, \f_{(s)}^{I_1}$. A more elegant way of dealing with this issue is to
construct gauge invariant combinations of the fluctuations. Such gauge invariant
combinations of the fluctuations were constructed in \cite{Skenderis:2006uy}, with the combinations reducing to the de Donder-Lorentz gauge fluctuations on
imposing this gauge.

In \cite{Skenderis:2006uy} the equations of motion satisfied by the fluctuations were constructed to quadratic order in the fluctuations, and the relation between five-dimensional fields
and ten-dimensional fields was also constructed up to quadratic order in the fluctuations. In the current context we are only interested in the five-dimensional Einstein metric
and we work only to linear order in the fluctuations. We can therefore read off from \cite{Kim:1985ez,Skenderis:2006uy} the relationship between the five-dimensional Einstein metric perturbation $h^{E}_{mn}$
and the ten-dimensional fields as
\be
h^{E}_{mn} = h^{0}_{mn} + \frac{1}{3} \pi^0 g_{mn}^o, \label{ein-def}
\ee
where the superscripts indicate that these are zero modes, i.e. associated with the trivial constant scalar harmonic. 

The type IIB supergravity equations lead to the linearized equation for the Einstein metric 
\be
({\cal L}_E + 4) h^E_{mn} = 0, 
\ee
where ${\cal L}_E$ is the Einstein operator, defined as usual by
\be
{\cal L}_E h^E_{mn} = \frac{1}{2} \left ( - \Box h^E_{mn} + D_p D_m h^{E p}_{n} + D_{p} D_{n} h^{E p}_{m} - D_{m} D_{n} h^{E p}_{p} \right ).
\ee 
The five-dimensional equation of motion in turn follows from reducing the ten-dimensional action 
\be
I_{IIB}  = \frac{1}{2 \kappa_{10}^2} \int d^{10}x \sqrt{-{\rm det}(g_{MN})} \left ( R(g_{MN}) - \frac{4}{5!} F_{MNPQR} F^{MNQR} + \cdots \right )
\ee
over the five-sphere\footnote{Note that the 10d action must be supplemented with a self-duality constraint for the five form field strength.}.
This results in
\be
I = \frac{{N}_c^2}{2 \pi^2} \int d^5 x \sqrt{-{\rm det} (g_{mn})} \left ( \frac{1}{4} R(g_{mn}) + \cdots \right ), 
\ee 
where we use the relation
\be
\frac{1}{2 \kappa_{10}^2} V_{S^5} = \frac{\pi^3}{2 \kappa_{10}^2} = \frac{1}{2 \kappa_5^2} = \frac{{N}_c^2}{8 \pi^2},
\ee 
which is applicable when the $AdS$ radius $L$ is set to one. Thus the effective Newton constant is given by
\be
\frac{1}{16 \pi G_{N}} = \frac{{N}_c^2}{8 \pi^2}. 
\ee

For the probe brane system, the type IIB supergravity equations are solved with source terms, from the D-brane action, which in turn implies that the linearized Einstein equations in five dimensions are sourced. The complete ten-dimensional action is
\be
I = I_{IIB} + I_{D7}
\ee
where
\be
I_{D7} = - T_7 \int d^{10}x  \int d^{8} \sigma \delta(x^M - X^M(\sigma^{\alpha})) e^{-\phi} \sqrt{- {\rm det} (g_{MN} \partial_\alpha X^M \partial_\beta X^N) + \cdots}.
\ee
Here $\sigma^{\alpha}$ denote the world-volume coordinates and the ellipses denote terms involving the world-volume gauge fields and Wess-Zumino couplings. The latter
do not contribute in the case of the D7-brane embeddings under consideration here. 

The source term results in a stress energy tensor \cite{Duff:1994an}
\be
T^{MN} = - T_7  \int d^8 \sigma \sqrt{-\gamma}e^{-\phi}  (\gamma^{\alpha \beta} \partial_\alpha X^M \partial_\beta X^N)  \frac{\delta(x^M - X^M(\sigma^{\alpha}))}{\sqrt{- {\rm det} (g_{MN})}}, 
\ee
where we denote the worldvolume induced metric as $\gamma_{\alpha \beta} = g_{MN} \partial_{\alpha} X^M \partial_{\beta} X^N$. 
The sourced IIB equation is thus
\be
(R_{MN} - \frac{1}{6} F_{MPQRS} F_{N}^{\; PQRS} + \cdots) = \kappa_{10}^2 (T_{MN} - \frac{1}{8} T g_{MN})\equiv \kappa_{10}^2 \bar{T}_{MN}, \label{iib-corr}
\ee
with $T = g^{MN} T_{MN}$. The trace adjusted stress energy tensor can be expanded in $S^5$ harmonics, using the same harmonic basis as for the metric:
\bea \label{T}
\bar{T}_{m n}(x,y) &=& \sum \bar{T}^{I_1}_{m n}(x) Y^{I_1}(y)  \\
\bar{T}_{m a} (x,y)&=& 
\sum (\tilde{T}^{I_5}_{(v)m}(x) Y_a^{I_5}(y) 
+ \tilde{T}^{I_1}_{(s) m}(x) D_a Y^{I_1}(y)) \nonumber \\
\bar{T}_{(ab)}(x,y) 
&=& \sum \bar{T}_{(t)}^{I_{14}}(x) Y_{(ab)}^{I_{14}}(y) 
+ \bar{T}^{I_5}_{(v)}(x) D_{(a} Y^{I_5}_{b)}(y)
+ \bar{T}^{I_1}_{(s)}(x) D_{(a} D_{b)} Y^{I_1}(y) )\nonumber \\
\bar{T}_{a}^a(x,y) &=& \sum \tilde{T}^{I_1}(x) Y^{I_1}(y) \nonumber
\eea
The correction to the five-dimensional Einstein equation only depends on the following zero modes (see Appendix A):
\be
({\cal L}_E + 4) h^E_{mn} =  \kappa_{10}^2 (\bar{T}^{0}_{m n} + \frac{1}{3} \tilde{T}^0 g^o_{mn}) \equiv  \bar{t}_{mn}, \label{src1}
\ee
where we have used the fact that the D7-brane embedding of interest does not source the RR five-form field strength. 

Given the D7-brane embedding, i.e. $ \theta (z) = \cos^{-1}(m z)$, the ten-dimensional energy momentum tensor source can be computed as
\be
T_{MN} = - T_7 {\cal T}_{MN} \delta (\theta - \theta(z)) \delta (\phi),
\ee
with
\bea
{\cal T}_{zz} &=& \frac{1}{z^2} (1-m^2 z^2)^2; \qquad {\cal T}_{\mu \nu} = \frac{1}{z^2} (1 -m^2 z^2) \eta_{\mu \nu}; \\
{\cal T}_{\Omega_3} &=& (1 - m^2 z^2)^2 g_{\Omega_3}; \qquad {\cal T}_{\phi \phi} = 0; \nn \\
{\cal T}_{\theta \theta} &=& m^2 z^2 (1 - m^2 z^2). \nn
\eea
where note that ${\cal T} \equiv g^{MN} {\cal T}_{MN}=8(1-m^2z^2)$. Here we denote the metric on the unit three-sphere by $g_{\Omega_3}$. 
The energy momentum tensor source can be projected onto spherical harmonics using Fourier decompositions of the delta functions:
\bea
\delta ( \theta - \theta(z)) &=& \frac{2}{\pi} + \sum_{m=1}^{\infty} \frac{4}{\pi} \cos (m \theta(z)) \cos (m \theta); \\
\delta (\phi) &=& \frac{1}{2\pi} + \sum_{m=1}^{\infty} \frac{1}{\pi} \cos (m \phi). \nn
\eea
By projecting onto the zero mode, one can then immediately show that the ten-dimensional energy momentum tensor source is such that
\be
\bar{T}^0_{zz} = - T_7 \frac{m^2}{\pi^2} (1 - m^2 z^2); \qquad \bar{T}^0_{\mu \nu} = 0,
\ee
with
\be
\bar{T}^0 = \frac{T_7}{\pi^2} (1 - m^2 z^2) (m^2 z^2 - 2),
\ee
and hence using \eqref{src1} we find that 
\be
\bar{t}_{zz} = - \frac{ t_0 }{3  z^2}(1 - m^4 z^4); \qquad
\bar{t}_{\mu \nu} = \frac{t_0}{6  z^2}(1-m^2 z^2)(m^2 z^2  -2) \eta_{\mu \nu},
\ee
where as before
\be
t_{0} = 16 \pi G_N (2 \pi^2 T_7). 
\ee
Given the five-dimensional stress tensor $t_{mn} = \bar{t}_{mn} - \frac{1}{2} \bar{t} g^o_{mn}$, it is straightforward to see that the perturbation of the Einstein metric induced by this source is in agreement with that given in \eqref{metric-p} and \eqref{g-inv}.

\subsection{Generalizations to other probe brane systems}

For a general probe brane system, the complete ten-dimensional action is
\be
I = I_{IIB} + I_{Dp}
\ee
where
\bea
I_{Dp} &=& - T_p \int d^{10}x  \int d^{p+1} \sigma \delta(x^M - X^M(\sigma^{\alpha})) e^{-\phi} \sqrt{- {\rm det} (\gamma_{\alpha \beta} + {\cal F}_{\alpha \beta})} \\
&& \qquad + T_p \int d^{10}x  \int  \delta(x^M - X^M(\sigma^{\alpha})) \left [ e^{\cal F} \wedge \sum_q C_q \right ],  \nn 
\eea
with 
\bea
\gamma_{\alpha \beta} &=& g_{MN} \partial_{\alpha} X^{M} \partial_{\beta} X^{N}; \\
{\cal F}_{\alpha \beta} &=& B_{MN} \partial_{\alpha} X^{M} \partial_{\beta} X^{N} + {F}_{\alpha \beta}; \nn \\
C_{\alpha_1 \cdots \alpha_{q}} &=& C_{M_1 \cdots M_1} \partial_{\alpha_1} X^{M_1} \cdots \partial_{\alpha_q} X^{M_q}. \nn
\eea
We consider embedding a brane into a type IIB background which is either $AdS_5 \times S^5$ or $AdS_5$ Schwarzschild  $\times \hspace{1mm}S^5$, so that the only background field profiles are for the metric and the five-form. Following the arguments in the previous section, we therefore only need to consider the equations for the metric and five-form perturbations, as the other perturbation equations decouple. 

The energy momentum tensor source is \cite{Duff:1994an,Skenderis:2002vf}
\be
T^{MN} = - T_p  \int d^{p+1} \sigma \sqrt{- M}e^{-\phi}  (M^{\alpha \beta} \partial_\alpha X^M \partial_\beta X^N)  \frac{\delta(x^M - X^M(\sigma^{\alpha}))}{\sqrt{- {\rm det} (g_{MN})}}, 
\ee
where we define  $M_{\alpha \beta} = \gamma_{\alpha \beta} + {\cal F}_{\alpha \beta}$ with $M^{\alpha \beta}$ being its inverse. 
The source in the five form equation of motion is 
\bea
&&\partial_{M} \left ( \sqrt{-g} F^{M NPQR} \right ) = \\
 &&  \qquad 2 \kappa_{10}^2 T_p \int d^{p+1} \sigma \delta(x^M - X^M(\sigma^{\alpha})) 
\epsilon^{\alpha_1 \cdots \alpha_{p+1}} {\cal F}_{\alpha_1 \alpha_2} \cdots \partial_{\alpha_{p-2}}X^{N} \partial_{\alpha_{p-1}} X^{P} \partial_{\alpha_{p}} X^Q \partial_{\alpha_{p+1}} X^R \nn 
\eea
Note that for a D7-brane this term only contributes if ${\cal F} \wedge {\cal F} \neq 0$. Therefore, provided that ${\cal F} \wedge {\cal F} = 0$, 
the correction to the five-dimensional Einstein equation due to source D7-branes still depends only on the stress energy tensor zero modes:
\be
({\cal L}_E + 4) h^E_{mn} =  \kappa_{10}^2 (\bar{T}^{0}_{m n} + \frac{1}{3} \bar{T}^0 g^o_{mn}) \equiv  \bar{t}_{mn}. \label{src2}
\ee
One can thus compute the perturbation to the five-dimensional Einstein metric by projecting the brane energy momentum source onto the appropriate combination of (spherical) zero modes. It would be straightforward to relax the condition ${\cal F} \wedge {\cal F} = 0$ and obtain the correction to the five-dimensional Einstein equation, taking into account the sources in the RR field equations, but we will not analyse this case in detail here. 

\bigskip

The analysis above immediately allows us to treat D7-branes at finite mass, density (and temperature). The finite temperature background can be written as
\bea
ds^2 &=&\rho^2  \left [ - \frac{f^2}{\tilde{f}} dt^2 + \tilde{f} dx^2 \right ] + \frac{d \rho^2}{\rho^2} + d \Omega_5^2; \label{finiteTbackground}\\
f(\rho) &=& 1 - \frac{u_0^4}{\rho^4}; \qquad \tilde{f}(\rho) = 1 + \frac{u_0^4}{\rho^4}. \nn 
\eea
with the temperature being $T = \sqrt{2} u_0/\pi$. The D7-brane embeddings can be expressed in terms of two scalar functions $\chi(\rho)$ and $a(\rho)$:
\be
\theta (\rho) = \cos^{-1} (\chi(\rho)); \qquad F_{\rho t} = \partial_{\rho} a(\rho) \equiv E(\rho), \label{embeddings}
\ee
where the potential is $A_t = a(\rho)$. These embeddings can be found numerically, see \cite{Mateos:2006nu,Kobayashi:2006sb}. The main feature is that at any finite density, i.e. whenever the asymptotic form
of the potential is
\be
A_t = \mu - \frac{\tilde{d}}{\rho^2} + \cdots \label{gaugeasymp}
\ee
with non-zero charge density $\tilde{d}$, the embeddings do not close off at finite radius. At finite temperature the embeddings have a spike which extends into the horizon, while at zero temperature this spike passes through the Poincar\'{e} horizon. In other words, asymptotically as $\rho \rightarrow \infty$, the brane wraps the equator $\theta = \pi/2$ of the five sphere but there is a spike, $\theta \rightarrow 0$ as $\rho \rightarrow u_0$. In the zero temperature limit, the spike solution becomes analytic for $\rho \rightarrow 0$:
\be
E(\rho) \approx {\cal E}; \qquad 
\theta(\rho) \approx {\cal \theta}_1 \rho, \label{spike}
\ee 
with ${\cal E}$ and $\theta_1$ constant. 

Focussing on the zero temperature limit for simplicity, the effective source stress energy tensor of \eqref{src2} is given in terms of $\theta(\rho)$ and $E(\rho)$ by
\bea
\bar{t}_{\rho \rho } &=&- t_{0} \frac{\sin^3 \theta}{2 \rho^2 (1  + \rho^2 \dot{\theta}^2 - E^2)^{\frac{1}{2}}} \left ( - \frac{2}{3} - \frac{4 \rho^2 \dot{\theta}^2}{3} + E^2 \right ); \\
\bar{t}_{tt} &=&  t_0 \frac{\rho^2 \sin^3 \theta}{2 (1  + \rho^2 \dot{\theta}^2 - E^2)^{\frac{1}{2}}} \left ( - \frac{2}{3} - \frac{ \rho^2 \dot{\theta}^2}{3} + E^2 \right ); \nn \\
\bar{t}_{ij} &=&  - t_0 \frac{\rho^2 \sin^3 \theta}{2 (1  + \rho^2 \dot{\theta}^2 - E^2)^{\frac{1}{2}}} \left ( - \frac{2}{3} - \frac{ \rho^2 \dot{\theta}^2}{3} \right ) \delta_{ij}. \nn
\eea
Here $\dot{\theta}$ denotes $\partial_{\rho} \theta(\rho)$. To compare with the previous sections we change coordinates to $z = 1/\rho$, and use the five-dimensional stress tensor $t_{mn} = \bar{t}_{mn} - \frac{1}{2} \bar{t} g^o_{mn}$:
\bea
{t}_{zz} &=& - t_{0} \frac{\sin^3 \theta}{2 z^2 (1  + z^2 {\theta}_z^2 - E^2)^{\frac{1}{2}}}; \label{5dstresstensordensity} \\
{t}_{tt} &=&  t_0 \frac{z^2 \sin^3 \theta}{2 (1  + z^2 \theta_z^2 - E^2)^{\frac{1}{2}}} \left ( 1+ z^2 {\theta}_z^2 \right ); \nn \\
{t}_{ij} &=&  - t_0 \frac{z^2 \sin^3 \theta}{2 (1  + z^2 {\theta}_z^2 - E^2)^{\frac{1}{2}}} \left ( 1 +  z^2 {\theta}_z^2 - E^2 \right ) \delta_{ij}, \nn
\eea
where $\theta_z = \partial_z \theta$. 

The metric perturbation induced by such sources can then be expressed as
\be
\delta (ds^2) = \frac{f(z)}{z^2} dz^2 - \frac{g(z)}{z^2} dt^2 + \frac{h(z)}{z^2} dx^i dx_i. \label{met-per2}
\ee
As previously the gauge invariant combination is
\be
\tilde{f}(z) = f(z) + z h'(z)
\ee
and there are now two independent Einstein equations:
\bea
- \tilde{f} - \frac{1}{4} z (g' - h') &=& \frac{1}{6} z^2 t_{zz}; \\
\frac{3}{2} z (h' - g') + \frac{1}{2} z^2 (g'' - h'') &=& z^2 (t_{tt} + t_{i}), \nn
\eea
where we define $t_{ij} = t_{i} \delta_{ij}$. These equations can be integrated to give
\bea
\tilde{f} (z) &=& - \frac{1}{6} z^2 t_{zz} - \frac{1}{2} z^4 \int \frac{dz}{z^3} \left (t_{tt} + t_i \right );  \label{metricpertfinitedensity}\\
(g(z) - h(z)) &=& 2 \int^z d\tilde{z} \tilde{z}^3 \int^{\tilde{z}} \frac{dw}{w^3} \left (t_{tt} (w) + t_i (w) \right ). \nn
\eea
These equations can be solved analytically as $z \rightarrow 0$ and $z \rightarrow \infty$. The near boundary expansions of
the fields $\chi$ and $E$ are
\bea
\chi &=& m z + c z^3 + \cdots \label{chiasymp} \\
E &=&  2 \tilde{d} z^3 + \cdots, \nn 
\eea
where $m$ is the quark mass, $c$ determines the quark condensate and $\tilde{d}$ is the density. The corresponding asymptotic expansions of the gauge invariant metric perturbations are
\bea
\tilde{f}(z) &=& \frac{t_0}{12} \left (1 - 2m^2 z^2 + { \cal O} \left ( m^4 z^4, m c z^4 \right ) + \cdots \right ); \label{asy-ex1}  \\
\left (g(z) - h(z)  \right ) &=& \frac{t_0}{3} \tilde{d} z^6 + \cdots \nn
\eea
Note that in fixing a Fefferman-Graham gauge as $z \rightarrow 0$ one needs to take into account the shift in the AdS radius. The Fefferman-Graham gauge is obtained by choosing
\be
f(z) = \frac{t_0}{12},
\ee
which then implies that 
\be
z h'(z) = \frac{t_0}{6}  \left (- m^2 z^2 + { \cal O} \left ( m^4 z^4, m c z^4 \right ) + \cdots \right ),
\ee
and hence
\be
h(z) = \frac{t_0}{12} \left (1 - m^2 z^2 + { \cal O} \left ( m^4 z^4, m c z^4 \right ) + \cdots \right ),
\ee
where the integration constant is fixed by the AdS radius. 

In the opposite limit of $z \rightarrow \infty$ we can use the spike solution \eqref{spike} to show that 
\bea
\tilde{f} (z) &=& \frac{t_0 \theta_1^3}{4  z^3 (1 - {\cal E}^2)^{\frac{1}{2}}}\left ( \frac{1}{3} + \frac{ {\cal E}^2}{7} \right ); \\
\left (g(z) - h(z) \right ) &=& - \frac{t_0 \theta_1^3 {\cal E}^2}{42 z^3 (1 - {\cal E}^2)^{\frac{1}{2}}}, \nn
\eea
and hence the metric perturbations are bounded in the deep interior. 

\bigskip

The effect of the metric perturbation \eqref{met-per2} on the entanglement entropy is expressed in exactly the same way as in previous sections, since $g(z)$ does not enter the entanglement entropy.  Thus for a slab, following \eqref{final-ee},
the brane contribution to the entanglement entropy is 
\begin{equation}
\delta S= \frac{L^2}{4 G_N \tilde{z}^2} \left(\int_a^b ds \frac{\sqrt{1-s^6}}{s^3} \tilde{f}(\tilde{z}s)- \left[h(\tilde{z}s)\frac{\sqrt{1-s^6}}{s^2} \right]^{s=b}_{s=a}\right) 
\end{equation}
where $\tilde{z}$ is the turning point of the original minimal surface and $a=\epsilon/\tilde{z}$. At zero density $\tilde{f}(z)$ is zero for $z \ge 1/m$ and continuity of the metric and its derivatives requires $h(1/m) = h'(1/m)= 0$. At any finite density $\tilde{f}(z)$ is non-zero at finite $z$ and there is no need to impose that the function $h(z)$ vanishes at a finite value of $z$. In the finite density case the integration is therefore over the entire entangling surface, i.e. the upper limit $b =1$ and
\begin{equation}
\delta S=  \frac{L^2}{4 G_{N} \tilde{z}^2} \left(\int_a^1 ds \frac{\sqrt{1-s^6}}{s^3} \tilde{f}(\tilde{z}s) + \left[h( \e )\frac{\sqrt{1-a^6}}{a^2} \right] \right),
\end{equation}
since the other boundary term vanishes at $s =1$. 

Note that $\delta S$ has no discontinuities in its derivatives with respect to mass or to the width of the slab at finite density since $\tilde{f}(z)$ has no discontinuities in its derivatives at finite density. This provides another reason for viewing as unphysical the discontinuities discussed earlier. 

It is useful to define the difference between the entanglement entropy at finite density and that at zero density, for the same mass:
\be
\delta S - \delta S_{\tilde{d}=0} = \frac{L^2}{4 G_{N} \tilde{z}^2} \left(\int_a^1 ds \frac{\sqrt{1-s^6}}{s^3} \delta \tilde{f}(\tilde{z}s) + \left[\delta h( \e )\frac{\sqrt{1-a^6}}{a^2} \right] \right), \label{diff-ee}
\ee
where
\bea
\delta \tilde{f}(z) &=& \tilde{f}(z) - \frac{t_0}{12} (1 - m^2 z^2)^2; \label{defdeltaf} \\
\delta h(z ) &=& h(z) - h(z)_{\tilde{d}=0}, \nn
\eea
where we have used the analytic expression for $\tilde{f}(z)$ at zero density. From the asymptotic expansions \eqref{asy-ex1}, we can infer that the asymptotic expansion of $\delta \tilde{f}$ is
\be
\delta \tilde{f}(z) = t_{0} {\cal O}(m c z^4)  + \cdots \label{asymf}
\ee
We can also always choose a gauge such that 
\be
\delta h(z) = t_0 {\cal O} (z^4) + \cdots;
\ee
this simply corresponds to matching the gauge asymptotically at zero and finite density. Substituting into \eqref{diff-ee} the difference between the entanglement entropy at finite density and that at zero density is UV finite\footnote{Earlier discussions of the UV finiteness of terms in the entanglement entropy induced by a chemical potential may be found in \cite{Wolf:2006zzb,Gioev:2006zz,Cramer:2006fu}.} and only the integrated term contributes:
\be
\delta S - \delta S_{\tilde{d}=0} = \frac{L^2}{4 G_{N} \tilde{z}^2} \int_a^1 ds \frac{\sqrt{1-s^6}}{s^3} \delta \tilde{f}(\tilde{z}s). \label{subtractedEE}
\ee
Note however that this quantity does have discontinuities in its derivatives at $\tilde{z} = 1/m$, since the zero density quantity has such discontinuities.

\subsection{Numerical calculation of the entanglement entropy at finite density}
We now consider in detail the embeddings in (\ref{embeddings}) with the aim to explicitily carry out the computation of the flavor entanglement entropy in the case of finite density. We follow the analysis in \cite{Mateos:2006nu,Kobayashi:2006sb}, though unlike the latter we focus on the zero temperature case. The background, instead of \eqref{finiteTbackground}, is therefore $AdS_5 \times S^5$
\begin{equation}
ds^2=\rho^2[-dt^2+dx_3^2]+\frac{d\rho^2}{\rho^2}+d\theta^2+\textrm{sin}^2\theta d\Omega_3^2 + \textrm{cos}^2\theta d\phi^2
\end{equation}
and the probe D7-brane extends in $\{t,x_3,\rho,\Omega_3\}$. We consider an embedding $\theta(\rho)$, and in addition we introduce a $U(1)$ gauge field $A_t(\rho)$ on the worldvolume of the D7-brane in order to study the gauge theory at finite density and chemical potential. The DBI action for this probe brane then evaluates to 
\bea
I_{D7} = -T_7 \int d^8 \sigma \frac{\rho^3}{4}(1-\chi^2)\sqrt{1-\chi^2+\rho^2 (\partial_{\rho}\chi)^2-2 (1-\chi^2)F_{\rho t}^2}
\eea
where $\chi(\rho)\equiv\textrm{cos}\left[\theta(\rho)\right]$ and $F_{\rho t} (\rho)=\partial_{\rho}A_t(\rho)$ is the electric field. The equation of motion for the gauge field has solutions with asymptotics given by (\ref{gaugeasymp}), and since $I_{D7}$ does not depend explicitly on $A_t$, there is a constant of motion $d\equiv \delta I_{D7}/\delta F_{\rho t}$. 

For solving the resulting equations of motion it is useful to eliminate the gauge field $A_t$ from the action by performing a Legendre transform with respect to $d$. The equation of motion for $\chi$ can then be obtained from the Legendre transformed action $\tilde{I}_{D7}$ as 
\bea
\begin{aligned}
\partial_{\rho} \left[ \frac{\rho^5 (1-\chi^2)\dot{\chi}}{\sqrt{1-\chi^2+\rho^2 \dot{\chi}^2}} \sqrt{1+\frac{8 \tilde{d}^2}{\rho^6 (1-\chi^2)^3}}\right]= \hspace{65mm} &\\
-\frac{\rho^3 \chi}{\sqrt{1-\chi^2+\rho^2 \dot{\chi}^2}}\sqrt{1+\frac{8 \tilde{d}^2}{\rho^6 (1-\chi^2)^3}}\left[3(1-\chi^2)+2\rho^2 \dot{\chi}^2-24\tilde{d}^2\frac{1-\chi^2+\rho^2 \dot{\chi}^2}{\rho^6(1-\chi^2)^3+8\tilde{d}^2}\right] \label{chieqn}
\end{aligned}
\eea
where $\dot{\chi}\equiv \partial_{\rho}\chi$ and $\tilde{d}\equiv d/T_{D7}$.  It is straightforward to show that asymptotically solutions to this equation take the form given in \eqref{chiasymp}. 

We solve (\ref{chieqn}) numerically for a given $\tilde{d}$ with regular boundary conditions imposed in the deep interior. Recalling that $\chi = \cos \theta$ the spike solution is such that $\chi(0) \rightarrow 1$.
However, since $\dot{\chi}(\rho)=-\sin \theta \dot{\theta}$, $\dot{\chi}(0) = 0$ and is independent of the value of 
$\dot{\theta}(0)$. Therefore we instead set boundary conditions at $\rho = \rho_0 \ll1$:
\be
\chi(\rho_0) = 1 - \frac{1}{2} \delta^2; \qquad  \dot{\chi} (\rho_0) = -\alpha, 
\ee
with $\alpha > 0$ and $\delta^2 \ll 1$.
These boundary conditions correspond to
\be
\theta(\rho_0) = \delta; \qquad 
\dot{\theta}(\rho_0) = \frac{\alpha}{\delta}. 
\ee
Such conditions are consistent with the spike solution $\theta = \theta_1 \rho + \cdots$ in  \eqref{spike} provided that
$\alpha \sim \delta^2/\rho_0$; if the latter condition is satisfied the solutions can be smoothly continued to $\rho =0$. 
These boundary conditions differ from those used in \cite{Mateos:2006nu,Kobayashi:2006sb} due to the fact that we work at zero temperature. Note that the quark mass can be extracted from the embedding using $\lim_{\rho \rightarrow \infty}( \rho \chi)$ due to (\ref{chiasymp}). 

The equations of motion for the gauge field are then given by Hamilton's equations with $\tilde{I}_{D7}$ as Hamiltonian, which reproduce the fact that $\tilde{d}$ is a constant, together with the equation 
\begin{equation}
\partial_{\rho} {A}_{t} = 2 \tilde{d} \frac{\sqrt{1-\chi^2+\rho^2 \dot{\chi}^2}}{\sqrt{(1-\chi^2)[\rho^6(1-\chi^2)^3+8\tilde{d}^2]}}. \label{gaugedereqn}
\end{equation}
This equation can be integrated to give
\begin{equation}
{A}_t(\rho) = 2 \tilde{d} \int_0^\rho d\rho' \frac{\sqrt{1-\chi^2+\rho'^2 \dot{\chi}^2}}{\sqrt{(1-\chi^2)[\rho'^6(1-\chi^2)^3+8\tilde{d}^2]}} \label{gaugeeqn}
\end{equation}
where we have set ${A}_t( \rho) \rightarrow 0$ as $\rho \rightarrow 0$. The chemical potential ${\mu}$ is then given by ${A}_t(\infty)$ in the previous expression. Once the embedding $\chi(\rho)$ has been found above, one can compute ${F}_{\rho t}\equiv\partial_{\rho}{A_t}$ and ${\mu}$ from (\ref{gaugedereqn}) and (\ref{gaugeeqn}) respectively. 
Note that the parameter $\tilde{d}$ indeed characterises the density since
\be
\lim_{\rho \rightarrow \infty}( \rho^3 \partial_{\rho} A_t) = 2 \tilde{d} ,
\ee
in agreement with \eqref{gaugeasymp}. 

\begin{figure}
\begin{center}
\setlength{\unitlength}{0.80mm}
\includegraphics*[trim =4.2cm 11.5cm 0.5cm 11cm,width=1.3\linewidth]{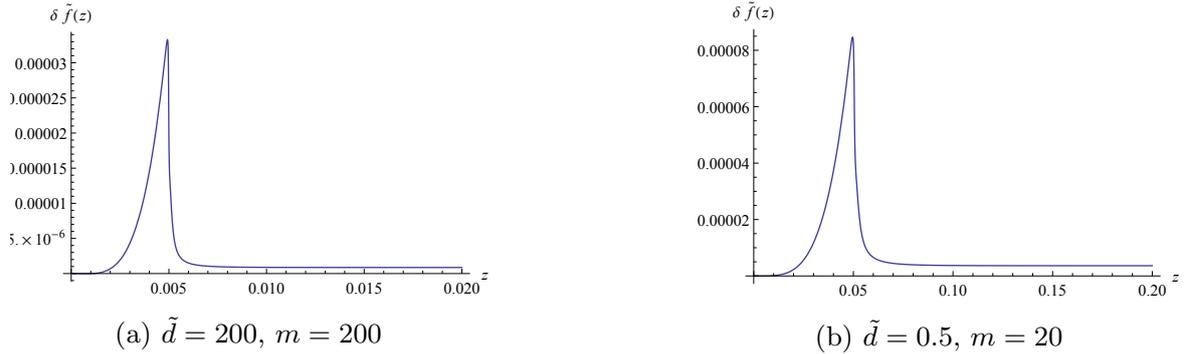}
\caption{Plots of $\delta \tilde{f}(z)$ for various values of $\tilde{d}$ and $m$. In all cases $\delta \tilde{f}(z)$ asymptotes to zero as $z \rightarrow \infty$.}
\label{deltaff}
\end{center}
\end{figure}

To compute the entanglement entropy one must change coordinates as in the previous section. At zero temperature the coordinate transformation is trivial and simply amounts to setting $z=1/\rho$ whilst leaving the other coordinates unchanged. One can then compute the five-dimensional stress tensor components in (\ref{5dstresstensordensity}), and thus the gauge-invariant metric perturbation $\tilde{f}(z)$ as defined by (\ref{metricpertfinitedensity}). We plot $\delta \tilde{f}(z)$, as defined in (\ref{defdeltaf}), in Figure \ref{deltaff} for various values of $\tilde{d}$ and $m$, the latter being fixed by the choice of both $\tilde{d}$ and $\chi'(0)$. The same general features are observed for all values of the parameters; $\delta \tilde{f}(z)$ peaks around $z=1/m$ and has a long spike slowly asymptoting to zero as $z \rightarrow \infty$. Although one might expect intuitively that the thickness of this spike is determined by the ratio $\tilde{d}/m$ (with a larger ratio leading to a thicker spike), the results indicate that it is in fact the magnitude of $\chi'(0)$ that determine this thickness, with a larger value of $\chi'(0)$ corresponding to a thicker spike (and a larger magnitude of  $\delta \tilde{f}(z)$ overall). 

\begin{figure}
\begin{center}
\setlength{\unitlength}{0.80mm}
\includegraphics[scale=0.5]{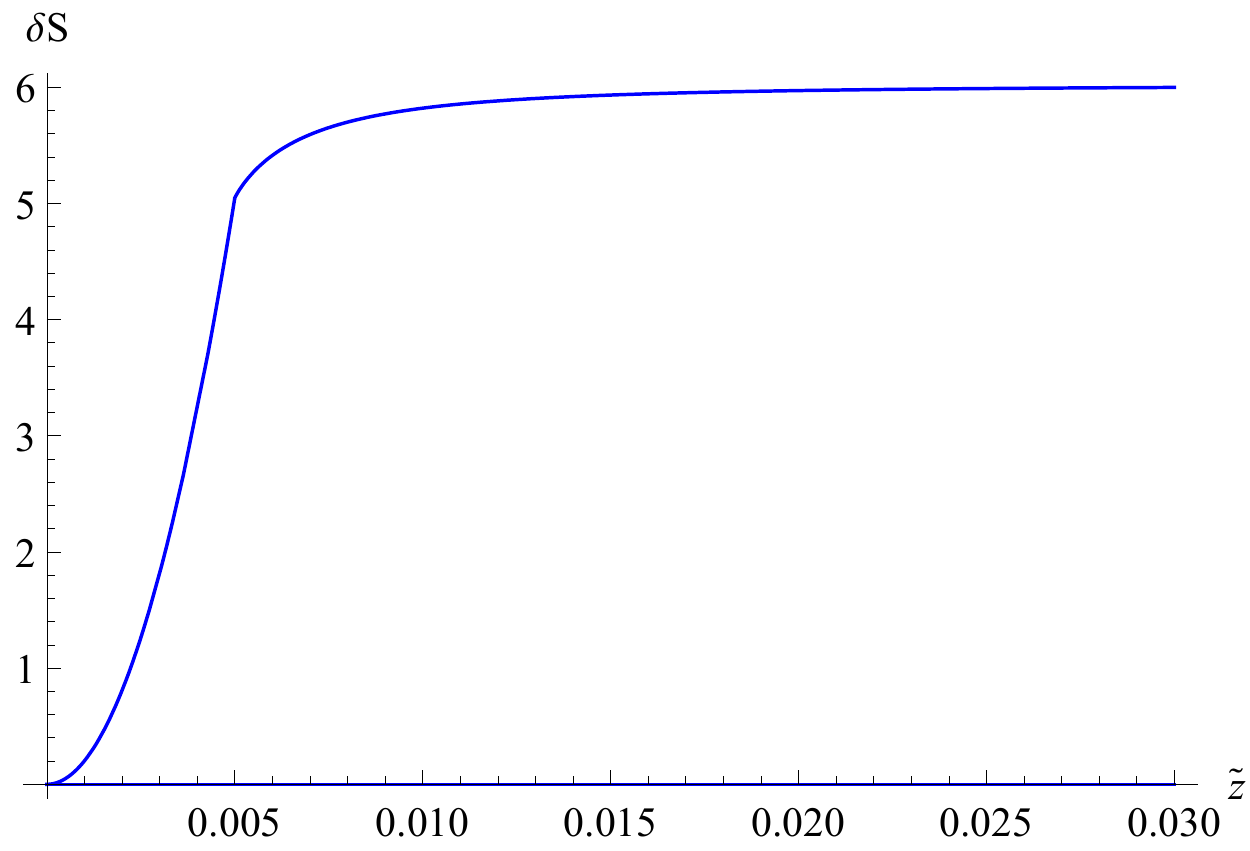}
\caption{Plot of the background subtracted entanglement entropy for $\tilde{d}=200$, $m=200$ as the width of the slab is increased. The entanglement entropy increases quadratically until the depth of the entangling surface is $1/m$ and then slowly saturates to a constant value as the width of the slab is increased further.}
\label{deltaS}
\end{center}
\end{figure}

It is a simple matter to compute the background subtracted entanglement entropy using (\ref{subtractedEE}): the result for $\tilde{d} = 200$, $m=200$ is shown in Figure \ref{deltaS}. The graph shows the entanglement entropy as a function of the 
depth of the entangling surface $\tilde{z}$, which is proportional to the slab width $l$. It follows from \eqref{asymf}  that the subtracted entanglement entropy increases quadratically with $\tilde{z}$ for $\tilde{z} \ll 1$; if $\delta f = \lambda z^4$ then 
\be
\delta S - \delta S_{\tilde{d}=0} = \frac{L^2\lambda }{48 G_N} \frac{\sqrt{\pi} \Gamma(1/3)}{\Gamma(11/6)} \tilde{z}^2.
\ee
The metric perturbation $\delta f$ reaches a maximum around $z \sim 1/m$ and is very small for $z > 1/m$, and therefore  the entanglement entropy of surfaces which extend to turning points $\tilde{z} \gg 1/m$ saturates. 

The D3-D7 system has a rich structure of phase transitions as the chemical potential, temperature and magnetic field are varied, see \cite{Evans:2010iy,Jensen:2010vd}. It would be interesting to use entanglement entropy to explore these phase transitions, extending the above results. Note that the entanglement entropy for massless flavors at finite density was discussed in \cite{Chang:2014oia}; our method would give the same results for massless flavors, and it would be interesting to explore how the entanglement entropy changes as one increases the ratio of density to mass. 

\section{Field theory interpretation} \label{six}

The D3-D7 system is dual to ${\cal N} = 4$ SYM  coupled to ${\cal N} =2$ massive hypermultiplets. The key features of this field theory are as follows. The field content of ${\cal N} = 4$ consists of the gauge fields $A_{\mu}$, scalars $X^A$ transforming in the fundamental of the R symmetry group $SO(6)$ and spinors $\lambda^i$ transforming in the spinor representation of $SO(6)$. The hypermultiplets consist of scalars $\chi$ and fermions $\eta$ transforming in the bifundamental of the $SU(N_c)$ and $SU(N_f)$ gauge groups. In the massless case the addition of these hypermultiplets preserves an $SO(4) \times SO(2)$ subgroup of the R symmetry group of ${\cal N} = 4$ SYM. The hypermultiplets are coupled to the ${\cal N} = 4 $ SYM fields by potential terms of the form
\be
I = \int d^4 x {\rm Tr}_{SU(N_c)} {\rm Tr}_{SU(N_f)} \left ( X^A \chi^{\dagger} \chi X^A \right )
\ee
Separating the branes by a distance $m$ (in string units) corresponds to introducing a mass term $m$ for the hypermultiplets, which breaks the conformal invariance and breaks the R symmetry group further to $SO(4)$.  

There are two distinct regimes of interest: the mass parameter $m$ being small relative to energy scales of interest, and the mass parameter being large compared to scales of interest. In the former case the theory is clearly described in terms of a small mass perturbation of a conformal field theory, and we can use the underlying conformal invariance to understand the entanglement entropy. Entanglement entropy for relevant perturbations has been studied recently \cite{Rosenhaus:2014zza} and we will discuss the relation to our results below. 

In the opposite regime, of high mass, 
at energy scales much lower than $m$ we can integrate out the hypermultiplets, effectively setting $\chi \sim \frac{1}{m}$. The potential term above controls the leading deformation to the ${\cal N} = 4$ SYM theory: at low energies the effective description must be 
\be
I = I_{SYM} + \frac{1}{m^2} \int d^4 x {\cal O}_{6}, \label{irr-def}
\ee
where ${\cal O}_6$ is an operator of dimension six in the ${\cal N} = 4$ SYM theory, i.e. it is an irrelevant deformation of the SYM conformal field theory. The dimension six operator explicitly breaks the R symmetry from $SO(6)$ to $SO(4)$ and is therefore charged with respect to the $SO(6)$ R symmetry of ${\cal N} = 4$ SYM. The backreaction of this deformation on the stress energy tensor, which is an R symmetry singlet, is necessarily quadratic in this deformation, i.e. the stress energy tensor is only affected at order $1/m^4$. The behaviour of entanglement entropy under irrelevant deformations has been less studied and we will explore this case in more detail below. 

\subsection{Zero mass: marginal deformation of CFT}

In the limit of zero mass, the brane contribution to the entanglement entropy of the slab is 
\be
\delta S = \frac{t_0L^2}{48 G_N} \left ( \frac{3}{2\e^2} + \frac{\sqrt{\pi}}{2 \tilde{z}^2} \frac{\Gamma(-1/3)}{\Gamma(1/6)} \right ),
\ee
where implicitly we have fixed a gauge choice such that 
\be
h(z) = \frac{t_0}{12}.
\ee
Note that this expression is proportional to the entanglement entropy of a slab in $AdS_5$:
\be
S = \frac{t_0L^2}{2 G_N} \left ( \frac{1}{2\e^2} + \frac{\sqrt{\pi}}{6 \tilde{z}^2} \frac{\Gamma(-1/3)}{\Gamma(1/6)} \right ). 
\ee
The entanglement entropy of a spherical surface in $AdS_5$ is
\be
S = \frac{\pi}{G_N} \left ( \frac{R^2}{2 \e^2} + \frac{1}{2} \log (\e/\e_{IR}) - \frac{1}{4} \right ).
\ee
The brane contribution to the entanglement entropy for a spherical surface at zero mass is 
\be
\delta S = \frac{t_{0} \pi}{8 G_N}\left ( \frac{R^2}{2 \e^2} + \frac{1}{2}  \log (\e/\e_{IR}) - \frac{1}{4} \right ),
\ee
which is again proportional to the AdS result. 

This is to be expected: suppose that the AdS radius is scaled as 
\be
L_{AdS} \rightarrow L_{AdS} \left (1 + \delta_{AdS} \right )
\ee
with $\delta_{AdS} \ll 1$. Since the bulk entangling surface is of dimension three, this implies that
\be
S \rightarrow S \left (1 + \delta_{AdS} \right )^{\frac{3}{2}} \approx S \left (1 + \frac{3}{2} \delta_{AdS} \right ). 
\ee
In the case at hand, the effect of the brane is to shift the AdS radius as
\be
\delta_{AdS} = \frac{t_0}{12}
\ee
The brane contribution to the entanglement entropy at zero mass is therefore precisely 
\be
\delta S =  \frac{3}{2} \delta_{AdS} S,
\ee
explaining the results above.

\subsection{Small mass: relevant deformation of CFT}

In the small mass regime, the mass $m$ is smaller than the energy scales of interest, so the system can be viewed as a mass perturbation of an underlying CFT. 

{\bf Half space:} Let us consider first the case in which the size of the slab $l \rightarrow \infty$, i.e. the space is divided into two regions by a plane; the brane contribution to the entanglement entropy is given in \eqref{infinitel}. In particular the logarithmic divergence is
\be
\delta S = \frac{t_0 L^2}{48 G_N} m^2 \textrm{log} (m \epsilon) = \frac{ \pi}{3} T_0 (m^2 L^2) \log (m \epsilon), \label{ee-log}
\ee
where in the latter expression we use \eqref{tens} i.e. $T_0$ is the effective tension of the D7-brane, reduced over the three sphere. 

The mass deformation in the field theory is associated with the following slipping mode of the D7-brane on the three sphere: letting the five sphere metric be 
\be
d\Omega_5^2 = d \theta^2 + \sin^2 \theta d \Omega_3^2 + \cos^2 \theta d \phi^2 
\ee
then the slipping mode is associated with the angle $\theta$, i.e. we retain only the following terms in the D7-brane action
\be
I = T_{0} \int d^{5} x \sqrt{g} \sin^3 \theta \sqrt{1 + g^{\mu \nu} \partial_{\mu} \theta \partial_{\nu} \theta},
\ee
where we have integrated over the three-sphere, because the mode of interest is an $SO(4)$ singlet and hence there is no dependence on the three sphere coordinates. The metric $g_{\mu \nu}$ denotes the $AdS_5$ metric and we work here in Euclidean signature as we need to compute correlation functions. 
The resulting equation of motion for $\theta$ is
\be
0 = \Box \theta - 3 \cot \theta - \frac{1}{2} \frac{ g^{\mu \nu} \partial_{\mu } \theta ( g^{\rho \sigma} \partial_{\rho} \theta \partial_{\sigma} \theta)}{1 + g^{\mu \nu} \partial_{\mu} \theta \partial_{\nu} \theta}
\ee
where $\Box$ is the Laplacian in the Euclidean $AdS_5$ metric. Linearising this equation around $\theta = \pi/2$ gives
\be
0 = \Box \theta + 3 \theta, \label{few}
\ee
i.e. the scalar is dual to an operator of dimension three. 

As we will discuss shortly, we are interested in computing the normalization of the two point function of this operator, and it thus suffices to consider the solution to the linearized equation of motion  \eqref{few}. From \cite{Karch:2005ms}, we can read off the operator one point function in terms of the asymptotic expansion of the scalar field:
\be
\langle {\cal O}_3 \rangle = T_0 \left ( - 2 \theta_{(2)} + \frac{1}{3} \theta_{(0)}^3 + \frac{R_0}{12} \theta_{(0)} + \Box_{(0)} \theta_{(0)} \right ) \label{onept}
\ee
where 
\be
\theta = z ( \theta_{(0)} + \theta_{(1)} z + \theta_{(2)} z^2  + \tilde{\theta}_{(2)} z^2 \log(z) + \cdots ) \label{thetaexp}
\ee
Here $\Box_{(0)}$ refers to the Laplacian in the boundary metric $g_{(0)}$ and $R_{(0)}$ is the scalar curvature of this metric, which is zero in our case. Working in momentum space, the regular solution to the linearized field equation \eqref{few} is
\be
\theta = \theta_{(0)} (k) (z^2 K_1 (k z)), \label{thetasol}
\ee
with $k$ the momentum and $\theta_{(0)} (k)$ corresponding to the Fourier transform of the source. One then expands \eqref{thetasol} about $z=0$ to identify the various terms in \eqref{thetaexp}, for example
\be
\theta_{(2)}(k) = \theta_{(0)}(k) \left( \frac{1}{4}(-1+2\gamma)k^2+\frac{1}{2}k^2 \textrm{log}\left( \frac{k}{2}\right)\right)
\ee
where $\gamma$ is the Euler constant. Functionally differentiating the one point function \eqref{onept} with respect to the source $\theta_{(0)}$, and setting the source to zero, therefore gives
\be
\langle {\cal O}_3(k) {\cal O}_3 (-k) \rangle = T_0 k^2 \log \left(\frac{k}{2}\right) + \cdots, 
\ee
where contact terms have been dropped. Fourier transforming back to position space gives 
\be
\langle {\cal O}_3 (x) {\cal O}_3 (0) \rangle = \frac{4 T_0}{\pi^2} {\cal R} \left ( \frac{1}{x^6} \right ), \label{holy}
\ee
where ${\cal R}$ denotes differential renormalization, see \cite{Skenderis:2002wp}.  

In \cite{Rosenhaus:2014zza} it was shown that for a CFT deformed by a relevant operator
\be
I \rightarrow I + \lambda \int d^{d} x {\cal O},
\ee
the change in the entanglement entropy of the half space is 
\be
\delta S = {\cal N} \lambda^2 \frac{(d-2)}{4 (d-1)} \frac{\pi^{\frac{d+2}{2}}}{\Gamma( \frac{d+2}{2})} {\cal A} \log \left (\frac{\epsilon_{UV}}{\epsilon_{IR}}  \right ), \label{smolkin}
\ee
where $\epsilon_{UV}$ and $\epsilon_{IR}$ correspond to UV and IR cutoffs, respectively, ${\cal A}$ is the area of the dividing surface while ${\cal N}$ is the normalisation of the two point function of ${\cal O}$, i.e. at separated points
\be
\langle {\cal O}(x) {\cal O}(0) \rangle = \frac{ {\cal N}}{x^{2 \Delta}},
\ee
with $\Delta = (d+2)/2$ the (relevant) operator dimension. Note that the entanglement entropy is unchanged to first order in the perturbation $\lambda$. 

In our case the normalisation is given in \eqref{holy}, $\lambda = m$ and $\mathcal{A}=L^2$. Hence
\be
\delta S = \frac{ \pi T_0}{3} m^2 L^2 \log \left (\frac{\epsilon_{UV}}{\epsilon_{IR}}  \right ).
\ee
which exactly agrees with \eqref{ee-log}, taking the IR cutoff to be $\e_{IR} = 1/m$. 

{\bf Sphere:} For the spherical entangling region the logarithmically divergent terms are
\be
\delta S = \frac{t_0 \pi}{8 G_N} \left  ( \frac{2 \mu^2}{3} + \frac{1}{2}  \right ) \log (m \e) = T_0 \left  ( \frac{4}{3} (\pi m R)^2 + \pi^2 \right ) \log (m \e). 
\ee
The second of these terms was explained above. 
The first is proportional to the mass deformation and can be expressed in the same form as \eqref{smolkin}, with 
\be
{\cal A} = 4 \pi R^2; \qquad
{\cal N} = \frac{4 T_0}{\pi^2}, 
\ee
setting the IR cutoff to be $\e_{IR} = 1/m$. 

{\bf Slab:} For a slab of finite width, the logarithmically divergent terms are precisely twice those for the half space:
\be
\delta S = \frac{ 2 \pi T_0}{3} m^2 L^2 \log \left (\frac{\epsilon_{UV}}{\epsilon_{IR}}  \right ).
\ee
Taking into account that the area of the entangling surface is in this case $2 L^2$, we again find exact agreement with \eqref{smolkin}. 

\bigskip

At first sight it may seem surprising that the expression \eqref{smolkin}, which was derived for the half space using the known modular Hamiltonian, is applicable
to other entangling geometries, with the entangling area replaced by the appropriate value. However, from the field theory perspective, one could derive the result
for the spherical region by conformal transformations (the modular Hamiltonian is also known) and the divergent contributions for a slab, being local, must necessarily give exactly twice the result for an infinite slab. It was argued in \cite{Rosenhaus:2014zza} that the result should hold for any geometry, since any entangling surface is locally flat; of course for a curved surface there are additional contributions to the entanglement entropy beyond this universal contribution. The dilaton effective action approach was also used in \cite{Banerjee:2014daa,Banerjee:2014hqa} to derive the logarithmic divergences for any shape entangling region, up to a universal coefficient computable from the dilaton effective action.

There is also a very simple holographic way to understand why the formula \eqref{smolkin} is applicable to the logarithmic divergences of any shape entangling surface, generalising the work of \cite{Hung:2011ta,Park:2015dia}. 
Deforming the conformal field theory by an relevant scalar operator corresponds in the bulk to coupling gravity to a massive scalar $\Phi$ i.e. we consider
\be
I = \frac{1}{16 \pi G_N} \int d^{d+1}x \left ( R + d(d+1) -  \frac{1}{2} (\partial \Phi)^2 - \frac{1}{2} M^2 \Phi^2 + \cdots  \right ),
\ee
where $M^2 = \Delta ( \Delta - d)$, with $\Delta <  d$ the dimension of the dual operator. Here implicitly we are working perturbatively in the scalar field so we include only quadratic terms in $\Phi$ with the ellipses denoting higher order terms. The normalisation of the operator two point function is \cite{Freedman:1998tz,Skenderis:2002wp}
\be
\langle {\cal O} (x) {\cal O} (0) \rangle = \frac{(2 \Delta -d) \Gamma(\Delta)}{16 \pi G_N \pi^{\frac{d}{2}} \Gamma (d - \frac{\Delta}{2})} \frac{1}{x^{2 \Delta}},
\ee
for $\Delta = d/2 + k$ with $k$ an integer. In particular we can write
\be
{\cal N} = \frac{\Gamma(d/2+1)}{8 \pi^{d/2+1} G_N}  \label{norm-2}
\ee
for $\Delta = d/2 + 1$. 

Working perturbatively around an $AdS_{d+1}$ background, a scalar field profile 
\be
\Phi = \lambda z^{d - \Delta}
\ee
corresponds to deforming the field theory by the dimension $\Delta$ operator, with $\lambda$ characterising the deformation. 
At quadratic order in the source there is a backreaction on the metric. Letting the metric perturbation be as before
\be
\delta (ds^2) = \frac{1}{z^2} \left ( f(z) dz^2 + h(z) dx^{\mu} dx_{\mu} \right )
\ee
then the Einstein equation implies that the gauge invariant combination of these perturbations is given by  
\be
f(z) + z h'(z) = \frac{(\Delta -d)}{2(d-1)} \lambda^2 z^{2 (4 -\Delta)}. 
\ee
Working in Fefferman-Graham gauge we may set $f(z) = 0$ in which case
\be
h(z) = - \frac{1}{4(d-1)} \lambda^2 z^{2 ( d - \Delta)} = - h_0 \lambda^2 z^{2 (d- \Delta)}.
\ee
Now consider an entangling surface in the deformed metric. Let the induced metric for the minimal surface be $\gamma_{ab} = \partial^{a} X^{\mu} \partial^b X^{\mu} g_{\mu \nu}$ 
and fix a static gauge such that 
\be
Z = z; \qquad X^{a} = \sigma^a; \qquad X^{i} = X^{i}(z,\sigma^a)
\ee
where asymptotically as $z \rightarrow 0$
\be
X^{i}(z, \sigma^a) = X^{i}(\sigma^a) + \cdots
\ee
The entanglement entropy contains divergent terms from
\be
S = \frac{1}{4 G_N} \int d^{d-1}x \sqrt{\gamma} \approx \frac{1}{4 G_N} \int dz d \sigma^a \sqrt{\gamma^o} \frac{1}{z^{d-1}} (1 + \cdots + \frac{(d-2)}{2} h(z) + \cdots)
\ee
where 
\be
\gamma^o_{ab} = \partial_a X^{i}(\sigma^c) \partial_b X^{j}(\sigma^c) \delta_{ij}
\ee
is the induced metric for the entangling surface on the boundary. Integrating over the radial coordinate one finds the usual power law volume divergence
\be
\frac{1}{4 G_N (d-2) \e^{d-2}} \int d \sigma^a \sqrt{\gamma^0} = \frac{\cal A}{4 G_N (d-2) \e^{d-2}},
\ee
with $z=\e$ being the UV cutoff and ${\cal A}$ being the volume of the entangling surface in the boundary. From the relevant perturbation one obtains a logarithmic divergence whenever
\be
\Delta = \frac{1}{2} (d+2)
\ee
where
\be
\delta S = \frac{{\cal A}}{8 G_N} (d-2) \lambda^2 h_0 \log(\e/\e_{IR}).
\ee
Using the identity \eqref{norm-2} we thence obtain
\be
\delta S = {\cal N} {\cal A} \lambda^2 \frac{(d-2) \pi^{d/2+1}}{4 (d-1)\Gamma(d/2+1)} \log(\e/\e_{IR}),
\ee
which is exactly \eqref{smolkin} but does not assume any geometry for the entangling surface. It would be interesting to prove this result from the field theory; the modular Hamiltonian for generic entangling surfaces is not known but the holographic result suggests that one should be able to compute the logarithmic divergences without complete knowledge of the modular Hamiltonian. 

Note that this result has a straightforward generalisation to irrelevant deformations. 
Deforming the conformal field theory by an irrelevant scalar operator corresponds in the bulk to coupling gravity to a massive scalar $\Phi$ with $M^2 = \Delta ( \Delta - d)$, with $\Delta > d$ the dimension of the dual operator. 
Following the same steps we see that the entanglement entropy contains divergent terms from
\be
S = \frac{1}{4 G_N} \int d^{d-1}x \sqrt{\gamma} \approx \frac{1}{4 G_N} \int dz d \sigma^a \sqrt{\gamma^o} \frac{1}{z^{d-1}} (1 + \frac{(d-2)}{2} h(z) + \cdots)
\ee
but since $\Delta > d$ the metric perturbation $h(z)$ always gives rise to additional UV divergences:
\be
 \delta S = - \frac{\lambda^2 (d-2)}{32 G_N (d-1)(2 \Delta - d - 2)} \frac{{\cal A}}{\e^{2 \Delta -d - 2}}. 
\ee
Using the identity \eqref{norm-2} we thence obtain
\be
\delta S = - {\cal N} \lambda^2 \frac{(d-2) \pi^{d/2+1}}{4 (d-1)(2 \Delta - d - 2) \Gamma(d/2+1)} \frac{{\cal A}}{\e^{2 \Delta -d - 2}}, \label{irrev}
\ee
where ${\cal N}$ is the operator normalisation. As for the usual power law divergences, such terms are not universal but nonetheless will be given an interpretation in the following section.

\subsection{Large mass: irrelevant deformation of CFT}

By large mass, we mean that the mass scale $m$ is higher than the energy scales of interest. This implies in particular that $m \gg 1/l$, where $l$ characterises the size of the entangling region, i.e. the width of the slab or the radius of the sphere. Thus we are always working in the regime $\mu \gg 1$. The UV divergent contributions to the entanglement entropy have already been explained above and here we are interested in explaining the leading finite contributions for $\mu \gg 1$. To decouple such contributions from the divergent terms, it is useful to look at the differentiated quantities \eqref{largue} and \eqref{sph4} (which  we argued previously do not receive contributions from UV divergent terms). 

As stated above, for large mass, the effective IR description is in terms of an irrelevant deformation of SYM.
It is easy to understand the effects of such a deformation on the entanglement entropy from the dual perspective. 
For a deformation by an operator of dimension six with $\lambda \sim 1/m^2$ the change in the metric behaves as $1/(mz)^4$. The metric perturbation can only be viewed as small relative to the background $AdS_5$ metric when $(mz)^4 \gg 1$. 

{\bf Slab:} Now it is straightforward to infer the effect of the irrelevant deformation on the entanglement entropy of a slab. Since the latter scales extensively with the volume of the slab, $L^2$, and the metric is corrected at order $1/m^4$, the change in the entanglement entropy goes as $L^2/m^4$. The underlying theory is conformal so the only scale in the problem is the width of the slab $l$. Since the entanglement entropy is dimensionless the effect of the massive modes is to change the entanglement entropy as
\be
\delta S \sim \frac{L^2}{m^4 l^6},
\ee
which indeed agrees with the term found in \eqref{largue}. 

We can infer this answer using \eqref{irrev},  which in the case of $\Delta = 6$ and $d =4$ gives
\be
\delta S \sim \lambda^2 \frac{ {\cal A}}{\e^6}.
\ee
In the case at hand $\lambda \sim 1/m^2$. The description in terms of an irrelevant deformation of SYM is only valid provided that we consider entangling surfaces for which $l \gg 1/m$. The effective cutoff should therefore be $\e \sim l \gg 1/m$ and hence we reproduce the formula above. 

{\bf Sphere:} For a spherical entangling surface the leading contribution to the differentiated entanglement entropy at large $\mu$ behaves as \eqref{sph4}
\be
\delta S \sim \frac{1}{m^2 R^2} + {\cal O}\left ( \frac{1}{m^4 R^4} \right ).
\ee
Thus although there is a $1/m^4$ term (as above) this is not the leading contribution. A simple way to understand the origin of this term is by exploiting the CHM map \cite{Casini:2011kv}. The entanglement entropy for the spherical region is then computed by computing the entropy in the mass deformed theory on a hyperbolic space. Since the fields are conformally coupled the action contains the terms
\be
I = \frac{1}{2} \int d^4 x \sqrt{-g}  \left (  (\partial \chi)^2 + m^2 \chi^2  + \frac{1}{6} {\cal R}\chi^2 \right ),
\ee
where $\chi$ is a hypermultiplet scalar and ${\cal R} \sim 1/R^2$ is the Ricci scalar. Now when we integrate out the hypermultiplets we obtain additional terms in  \eqref{irr-def}: setting $\chi \sim 1/m$ we obtain a contribution from the curvature coupling of order $1/(m^2 R^2)$, which is in agreement with the expression above.

\subsection{Conformal perturbation theory at higher orders}

The brane contributions to the entanglement entropy for the half space \eqref{infinitel} consist of only the divergent term arising from the shift in the AdS radius and the logarithmic divergence discussed above. Since the logarithmic divergence is expressed entirely in terms of the coefficient of the two point function for the dimension three (fermion mass) operator, this contribution to the entanglement entropy is trivially not renormalized relative to the weak coupling result: since the operator dimension is protected, there is such a contribution regardless of the coupling. The result \eqref{infinitel} only includes powers of $m$ up to $m^2$, which follows on dimensional grounds: the dimensionless 
entanglement entropy scales extensively with the slab area $L^2$ and therefore the only way that contributions at order $m^3$ or higher could arise in the entanglement entropy would be if the latter was IR divergent, i.e. the contributions would have to scale as 
\be
\delta S \sim L^2 m^k \Lambda_{IR}^{k-2}
\ee
where the IR cutoff $\Lambda_{IR} \gg 1$ (relevant perturbations cannot introduce UV power law divergences so an IR cutoff is the only possibility). However, the entanglement entropy 
is an IR safe quantity and therefore no such dependence on an IR cutoff should arise. 

From the perspective of conformal perturbation theory, it is not obvious that there are not contributions to the entanglement entropy from higher order terms, i.e. terms of order $m^3$ or higher. The change in the entanglement entropy is in general expressed as \cite{Rosenhaus:2014nha,Rosenhaus:2014ula,Rosenhaus:2014zza}
\be
\delta S = - m \langle {\cal O} K \rangle  + \frac{1}{2} m^2 \left (\langle {\cal O} K K \rangle - \langle {\cal O} {\cal O} \rangle \right ) + { \cal O}(m^3)
\ee
where ${\cal O}$ is the deforming operator and $K$ is the modular Hamiltonian. The first term vanishes by conformal invariance and the second term gives the logarithmic divergence. By the argument above, higher order terms (dependent on higher order correlation functions) must vanish and it would be interesting to show this explicitly. 

\section{Differential entropy} \label{seven}

\begin{figure}
\begin{center}
\setlength{\unitlength}{0.50mm}
\includegraphics*[width=0.4\linewidth]{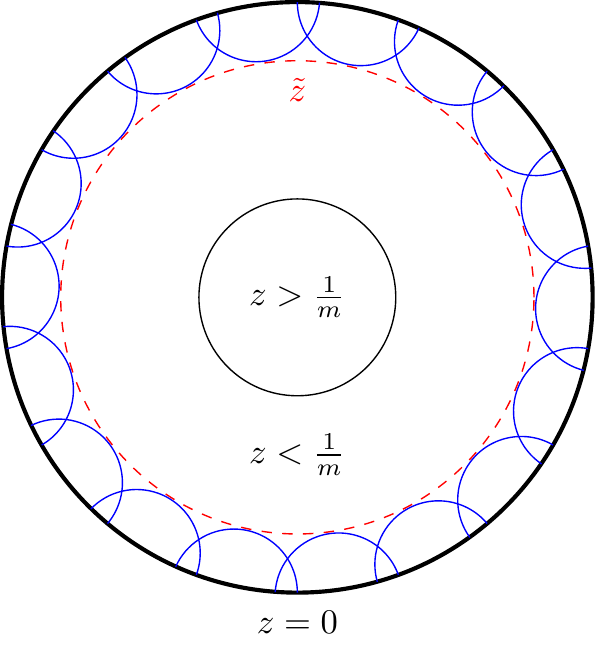}
\caption{An illustration of the equivalence between the differential entropy of a boundary partition and the area of a corresponding hole in the bulk. As the number of strips partitioning the boundary tends to infinity, the turning points of the associated minimal surfaces (blue) form a smooth hole in the bulk (red) whose area equals the differential entropy. }
\label{fig:diffentropy}
\end{center}
\end{figure}

Given the expression for the entanglement entropy of a slab, we now proceed to compute the differential entropy. Following \cite{Balasubramanian:2013rqa,Balasubramanian:2013lsa,Myers:2014jia,Balasubramanian:2014sra,Czech:2014wka,Headrick:2014eia}, the differential entropy should correspond to the area of a surface in the backreacted geometry - see Figure \ref{fig:diffentropy}.

The differential entropy is defined as 
\begin{equation}
E=\sum_{k=1}^{\infty}[ S(I_k)-S(I_k \cap I_{k+1})]
\end{equation}
where $\{ I_k\}$ is a set of intervals that partitions the boundary. We will cover the boundary with $n$ intersecting slabs -  we take $I_k$ to be a slab of width $\Delta x$, and the intersection $I_k \cap I_{k+1}$ is thus a strip of width $\Delta x - L_x/n$ where $L_x$ is the regularised length of the $x$-direction. At the end we will take the limit $n \rightarrow \infty$. 

For a slab in $AdS_5$ there is a relation between the strip width $\Delta x $, and the maximum bulk depth of the associated extremal surface $\tilde{z}$, see \eqref{rel1}. When the slab lies in the perturbed geometry this relation is modified (see for example \eqref{rel2}), and therefore it is useful to leave the relation implicit as 
\be 
\Delta x = c \tilde{z} = ( c_0 + c_1 t_0 + \mathcal{O}(t_0^2)) \tilde{z} . \label{c-def} 
\ee
with 
\be
c_0  = \frac{2 \sqrt{\pi} \Gamma ( \frac{2}{3})}{\Gamma ( \frac{1}{6})}; \label{c-def2} 
\ee
and $c_1$ depends on whether $\mu$ is greater than or less than one. The differential entropy takes the following form 
\begin{equation}
E = \lim_{n \to \infty} n[S(\Delta x)-S(\Delta x - L_x/n)] \label{de-def}
\end{equation}
where the overall factor of $n$ arises from the fact that all slabs are of equal width. For $AdS_5$ this gives
\be
E = \frac{L^2 \sqrt{\pi} \Gamma ( - \frac{1}{3}) }{12 \Gamma ( \frac{1}{6}) G_N}  \lim_{n \to \infty} n \left [ 
\frac{c^2}{(\Delta x)^2} - \frac{c^2}{(\Delta x - L_x/n)^2} \right ]
\ee
and hence 
\be
E = - \frac{ V \sqrt{\pi} \Gamma (- \frac{1}{3}) } {6 \Gamma (  \frac{1}{6}) G_N} \frac{c^2}{ (\Delta x)^3},
\ee
where $V = L^2 L_x$ is the regularised three-volume. Using \eqref{c-def2} this expression can be rewritten
as 
\be
E = \frac{V}{4 G_N} \frac{c_0^3}{(\Delta x)^3}, \label{e-ads5} 
\ee
and the latter is manifestly equal to the volume of the turning point surface (i.e. the hole) divided by $4 G_N$.

\subsection{Large mass $\mu \gg 1$}
Proceeding to calculate the flavor contributions, we begin with the very large mass case. The relevant expression for the entanglement entropy is \eqref{largem}.
The two leading terms are manifestly independent of $\tilde{z}$ and thus of $\Delta x$ which implies that the differential entropy vanishes in this limit:
\begin{equation}
\delta E = \lim_{n \to \infty} n[\delta S(\Delta x)-\delta S(\Delta x - L_x/n)] \rightarrow 0. 
\end{equation}
The limit $\mu \gg 1$ means $ \tilde{z} \gg 1/m$ and thus the hole formed in the bulk is far away from the probe brane (which ends at $z=1/m$). 
In the large $\mu$ limit the leading contribution to the differential entropy arises from the third term in \eqref{largem} and hence 
\bea
\delta E &=& - \frac{t_0 L^2}{2304 m^4 G_N}  \lim_{n \to \infty} n \left [ \frac{c^6}{\Delta x^6} - \frac{c^6}{(\Delta x - L_x/n)^6} \right ]; \label{de-pert} \\
&=& \frac{t_0 V c_0^6}{384 m^4 (\Delta x)^7 G_N}, \nn
\eea
where $c_0$ is defined by \eqref{c-def2}. Note that since the gauge dependent terms are independent of $\Delta x$ they also automatically cancel in the differential entropy. 

Now let us compare the differential entropy with the volume of a hole of radius $\tilde{z}$. Since the backreaction on the $AdS_5$ metric in this region is zero one might naively expect that the change in the differential entropy is zero. However, this does not take into account the fact that the relation between the turning point radius $\tilde{z}$ and the width of the boundary slabs $\Delta x$ is modified \eqref{rel2}. In other words, the gravitational entropy of the hole remains
\be
E_{grav} = \frac{V}{4 G_N} \frac{1}{\tilde{z}^3},
\ee
since the metric is $AdS_5$, but to express this quantity in terms of $\Delta x$ we need to use the relation in \eqref{c-def} to first order in the perturbation, resulting 
in 
\be
E_{grav} = \frac{V }{4 G_N} \left ( \frac{c_0^3}{ (\Delta x)^3} + \frac{3 c_0^2 c_1 t_0}{(\Delta x)^3} \right ) = \frac{V c_0^3}{4 G_N (\Delta x)^3} + \frac{t_0 V  c_0^6}{384 m^4 (\Delta x)^7 G_N}, \label{above2}
\ee
where we use
\be
c_1 =  \frac{1}{288 m^4 \tilde{z}^4} =  \frac{c_0^4}{288 m^4 (\Delta x)^4}.
\ee
where the latter equality is to order $t_0$. The second term in \eqref{above2} is in exact agreement with the calculation of the perturbation in the differential entropy \eqref{de-pert}; we will see in section 7.3 that this agreement holds for all $\mu > 1$.

\subsection{Small mass $\mu < 1$}

We proceed with the small mass case. Extracting from \eqref{smallmstrip} the terms which depend on $\tilde{z}$ one obtains
\begin{equation}
\delta S= \frac{t_0 L^2}{48 G_N} \left( \frac{\sqrt{\pi}}{12}\frac{\Gamma(-1/3)}{\Gamma(7/6)} \frac{1}{\tilde{z}^2} +  \frac{\sqrt{\pi}}{12}\frac{\Gamma(1/3)}{\Gamma(11/6)} m^4 \tilde{z}^2 - 2 m^2  \textrm{log}{\tilde{z}} + \cdots \right)
\end{equation}
From these terms in the entanglement entropy  we can now proceed to compute the differential entropy using \eqref{de-def}. Again it is clear that all divergent and gauge dependent terms cancel from the differential entropy, since they are independent of $\Delta x$. 
Noting that:
\begin{equation}
(\Delta x)^2 - (\Delta x-L_x/n)^2=2\Delta x \frac{L_x}{n}+\mathcal{O}(1/n^2)
\end{equation}
\begin{equation}
\frac{1}{(\Delta x)^2}-\frac{1}{(\Delta x -L_x/n)^2}=-\frac{2 }{\Delta x ^3}\frac{L_x}{n}+\mathcal{O}(1/n^2)
\end{equation}
\begin{equation}
\textrm{log}\frac{\Delta x - L_x/n}{\Delta x}=-\frac{1}{\Delta x }\frac{L_x}{n}+\mathcal{O}(1/n^2)
\end{equation}
we obtain the perturbation in the differential entropy:
\begin{equation}
\delta E = - \frac{t_0 V}{48G_N}\Bigg( - \frac{3 c_0^3}{2 \Delta x^3}-m^4\frac{\sqrt{\pi}}{6} \frac{\Gamma(1/3)}{\Gamma(11/6)}\frac{\Delta x}{c_0^2}+ \frac{2 m^2}{\Delta x} \Bigg), \label{small-m}
\end{equation}
where $c_0$ is as given in \eqref{c-def2}. 

Now let us compare this expression with the gravitational entropy of a hole of radius $z_t$:
\be
E_{grav}= \frac{V}{4 G_N} \frac{1}{z_t^3} \left ( 1 + \frac{3}{2} h(z_t) \right ),
\ee
where we take into account the metric perturbation relative to $AdS_5$. In this case the relation between $z_t$ and $\Delta$ is given by
\be
\Delta x = c_0 z_t ( 1 - \frac{1}{2} h(z_t)) + c'_1 t_0 z_t,
\ee
where $c_0$ is defined in \eqref{c-def2} and 
\be
c'_1 = - \frac{1}{24} \left ( - c_0  + \frac{4 \mu^2}{3} - \frac{\mu^4 \sqrt{\pi} \Gamma(\frac{1}{3}) }{9  \Gamma(\frac{11}{6})} \right )
\ee
Thus we note that the gauge dependent quantity $h(z_t)$ cancels from the entropy of the hole, with
\be
E_{grav} = \frac{V}{4 G_N} \frac{1}{ (\Delta x)^3} \left  (c_0^3 + 3 c_0^2 c_1' t_0 \right ),
\ee
and moreover the change in gravitational entropy of the hole agrees exactly with the change in the differential entropy \eqref{small-m}. 

\subsection{$\mu > 1$}

We now proceed with the case $\mu >1$, without requiring $\mu \gg 1$. The entanglement entropy for $\mu>1$ is given by equations (\ref{midm}) and (\ref{largem}):
\begin{equation}
\begin{split}
\delta S = \frac{t_0 L^2}{48 G \tilde{z}^2}\Big(\frac{1}{2a^2}+\frac{1}{6 \mu^4}{}_3 F_2\left(\{1/2,1,1\},\{2,2\},1/\mu^6\right)-\frac{\mu^2}{2}{}_2 F_1\left(-1/2,-1/3,2/3,1/\mu^6\right) \\+ \frac{\mu^2}{2}{}_2 F_1\left(-1/2,1/3,4/3,1/\mu^6\right) +2\mu^2 \textrm{log}(\mu a)\Big)+\delta S_{\textrm{scheme}}(m,\epsilon)
\end{split}
\end{equation}
Recalling that $\mu=m\tilde{z}$ and $a=\epsilon/\tilde{z}$, the divergent and log parts are independent of $\tilde{z}$ (and thus $\Delta x$) and so do not contribute to the differential entropy, nor do the gauge dependent terms. Also recall that $\tilde{z}=\Delta x /c$, where throughout the following we can replace $c \rightarrow c_0$ to order $t_0$ since $\delta S$ already contains an overall factor of $t_0$. We thus have:
\begin{equation}
\begin{split}
\delta S = \frac{t_0 L^2}{48 G}   \Big(\frac{c^6}{6 m^4 \Delta x^6} {}_3 F_2 
\left(\{1/2,1,1\},\{2,2\}, \frac{c^6}{m^6 \Delta x^6} \right) 
-\frac{m^2}{2}{}_2 F_1\left(-1/2,-1/3,2/3,\frac{c^6}{m^6 \Delta x^6}\right) \\+ \frac{m^2}{2}{}_2 F_1\left(-1/2,1/3,4/3,\frac{c^6}{m^6 \Delta x^6}\right) \Big) + ...
\end{split}
\end{equation}
where the ellipses denote $\tilde{z}$-independent terms. We want to compute the differential entropy:
\begin{equation}
\delta E = \lim_{n \to \infty} n[\delta S(\Delta x)-\delta S(\Delta x - L_x/n)]
\end{equation}
which requires the relations:
\begin{equation}
\begin{split}
{}_2 F_1\left(-1/2,1/3,4/3,\frac{a}{x^6}\right) - {}_2 F_1\left(-1/2,1/3,4/3,\frac{a}{(x-L_x/n)^6})\right)= \\ \frac{3 a L_x}{4 x^7 }{}_2 F_1\left(1/2,4/3,7/3,\frac{a}{x^6}\right)\frac{1}{n}+\mathcal{O}(1/n^2)
\end{split}
\end{equation}
\begin{equation}
\begin{split}
{}_2 F_1\left(-1/2,-1/3,2/3,\frac{a}{x^6}\right) - {}_2 F_1\left(-1/2,-1/3,2/3,\frac{a}{(x-L_x/n)^6})\right)= \\ -\frac{3 a L_x}{2 x^7 }{}_2 F_1\left(1/2,2/3,5/3,\frac{a}{x^6}\right)\frac{1}{n}+\mathcal{O}(1/n^2)
\end{split}
\end{equation}
\begin{equation}
\begin{split}
\frac{1}{x^6} {}_3 F_2 
\left(\{1/2,1,1\},\{2,2\}, \frac{a}{x^6} \right) -\frac{1}{(x-L_x/n)^6} {}_3 F_2 \left(\{1/2,1,1\},\{2,2\}, \frac{a}{( x-L_x/n)^6} \right)= \\ \frac{12 L_x}{a x}\left(-1+\sqrt{1-\frac{a}{x^6}}\right)\frac{1}{n} +\mathcal{O}(1/n^2)
\end{split}
\end{equation}
where here $a/x^6 <1$ since we are considering $\mu >1$. For the former two results we have used the standard identity:
\begin{equation}
\frac{d}{dz} {}_2 F_1 (a,b,c,x)=\frac{a b}{c} {}_2 F_1 (a+1,b+1,c+1,x)
\end{equation}
and for the latter result we have used the following relation:
\begin{equation}
 {}_3 F_2 
\left(\{1/2,1,1\},\{2,2\},x\right) =\frac{4}{x}\left(\textrm{log}\left(\frac{1+\sqrt{1-x}}{2}\right)-\sqrt{1-x}+1\right)
\end{equation}
where, again, here $x <1$ since we are considering $\mu>1$.
The differential entropy is then calculated to be:
\begin{equation}
\begin{split}
\delta E = \frac{t_0 V}{384 G}\frac{c_0^2}{(\Delta x)^3 \mu^4}\left( 16 \mu^6 \left( -1+\sqrt{1-\frac{1}{\mu^6}}\right)+ 6{}_2 F_1\left(\frac{1}{2},\frac{2}{3},\frac{5}{3},\frac{1}{\mu^6}\right)+3{}_2 F_1\left(\frac{1}{2},\frac{4}{3},\frac{7}{3},\frac{1}{\mu^6}\right) \right) 
\end{split}
\end{equation}
where recall $V \equiv L^2 L_x$, and we leave some $\Delta x$-dependence implicit in $\mu$ for notational clarity. 

We now want to compute the corresponding change in the gravitational entropy of the hole. As shown in section 7.1, the change in gravitational entropy of the hole is given by:
\begin{equation}
\delta E_{\textrm{grav}} = \frac{V}{4 G_N}\frac{3 c_0^2 c_1 t_0}{\Delta x^3}
\end{equation}
where $c_1$ is defined by equation (\ref{c-def}) and is given by equations (\ref{c1largem1})-(\ref{c1largem2}):
\begin{equation}
c_1=\frac{1}{288 \mu^4}\left( 16 \mu^6 \left( -1+\sqrt{1-\frac{1}{\mu^6}}\right)+ 6{}_2 F_1\left(\frac{1}{2},\frac{2}{3},\frac{5}{3},\frac{1}{\mu^6}\right)+3{}_2 F_1\left(\frac{1}{2},\frac{4}{3},\frac{7}{3},\frac{1}{\mu^6}\right) \right) 
\end{equation}
The equality of the differential entropy and the gravitational entropy of the hole is then manifest.

\section{Entanglement and differential entropy for top-down solutions} \label{eight}

The main focus of this work has been to develop a systematic method for computing the entanglement entropy for top-down brane probe systems. Our method however immediately generalises to any top-down solution which can be viewed as a perturbation of $AdS_5 \times S^5$ (or indeed a perturbation of the finite temperature $AdS_5$ Schwarzschild $\times \hspace{1mm} S^5$): using the Kaluza-Klein holography dictionary we can extract the effective five-dimensional Einstein metric and thence compute the entanglement entropy. 

It is worth noting that given a general asymptotically $AdS_5 \times S^5$ supergravity solution one cannot extract the five-dimensional Einstein metric, except asymptotically around the conformal boundary. Brane probe systems in which the perturbations relative to $AdS_5 \times S^5$ are small everywhere, and thus the Kaluza-Klein dictionary can be used to compute the five-dimensional Einstein metric at all scales, are special cases. In general the Kaluza-Klein dictionary becomes intractable when the metric perturbations relative to the $AdS_5 \times S^5$ background are of order one, i.e. at some finite distance from the conformal boundary. 

Another subset of ten-dimensional supergravity solutions can be expressed in terms of the uplifts of five-dimensional gauged supergravity solutions, for which the Einstein metric is known from the lower-dimensional theory. For example, specific cases of Coulomb branch solutions are realised as solutions of ${\cal N} = 8$ gauged supergravity in five dimensions; generic Coulomb branch solutions are however not realised as lower-dimensional solutions. The Coulomb branch examples also illustrate the fact that the causal structure of the five-dimensional Einstein metric is generically not the same as that of the uplifted ten-dimensional metric: the former can have naked timelike singularities which correspond to harmless null horizons in the uplifted solutions. One could thus envisage a scenario where the lower-dimensional metric had no entanglement shadows of the type discussed in \cite{Balasubramanian:2014sra} but the uplifted solution had shadow regions which could not be probed by five-dimensional fields at all. 

The entanglement entropy is computed from the five-dimensional Einstein metric and the differential entropy (built from entanglement entropy) therefore reconstructs areas of holes in the five-dimensional Einstein metric. The ten-dimensional metric cannot be reconstructed just from the five-dimensional Einstein metric: the uplift map requires all the matter fields in the lower-dimensional theory. Therefore the standard entanglement entropy and differential entropy cannot in principle reconstruct the ten-dimensional geometry without additional information.

\subsection{Coulomb branch examples}

To illustrate the above discussions we consider 
a particular Coulomb branch solution discussed in \cite{Freedman:1999gp,Freedman:1999gk}. The Coulomb branch of $\mathcal{N}=4$ SYM corresponds to the spontaneous breaking of the gauge symmetry by giving VEVs to the scalars - on the gravity side, these solutions are represented by multi-centre D3-brane solutions. These flows break superconformal invariance but preserve sixteen supercharges.

\cite{Freedman:1999gk} studies particular Coulomb branch solutions which admit consistent truncations. These flows are described in five-dimensional gauged supergravity by a single scalar field $\chi (r)$ where $r$ is a radial coordinate for the 5-dimensional metric
\begin{equation}
ds^2 = e^{2A(r)}dx_{\nu}dx^{\nu}+ dr^2 \label{5dgeo}
\end{equation}
where $r \rightarrow \infty$ corresponds to the conformal boundary. The BPS conditions can be written as
\begin{equation}
\frac{d \chi}{d r}=\frac{g}{2}\frac{\partial W}{\partial \chi} \hspace{10mm}\frac{dA}{d r}=-\frac{g}{3}W \label{kseqn}
\end{equation}
where $W(\chi)$ is the superpotential and $g$ is the gauged supergravity coupling constant. In particular, we will be interested in the solution which, from the ten-dimensional point of view, corresponds to the D3-branes being uniformly distributed on a disc of radius $\sigma$ in the transverse space, preserving $SO(4)\times SO(2)$ of the $SO(6)$ symmetry in the conformal $AdS_5 \times S^5$ solution. In this case one has 
\begin{equation}
W(\chi)=-e^{\frac{2}{\sqrt{6}}\chi}-\frac{1}{2}e^{-\frac{4}{\sqrt{6}}\chi} \label{superpot}
\end{equation}
\begin{equation}
A(\chi)=\frac{1}{2}\textrm{log}\left|\frac{e^{\frac{2}{\sqrt{6}}\chi}}{1-e^{\sqrt{6}\chi}}\right|+\textrm{log}\left(\sigma \right) \label{flow}
\end{equation}
where we have as usual set the radius of curvature of $AdS_5$ to be one. 

Redefining the radial coordinate one can write the metric as
\be
ds^2 = \lambda^2 \rho^2 \left ( dx_{\nu} dx^{\nu} +\frac{d\rho^2}{\rho^4 \lambda^6} \right ) \qquad
\lambda^6 = \left ( 1 + \frac{\sigma^2}{\rho^2} \right ), \label{ocs}
\ee
where $\rho \rightarrow \infty$ at the conformal boundary and the metric is $AdS_5$ for $\sigma = 0$. The uplifted ten-dimensional metric is then expressed as 
\be
ds_{10}^2 = \Delta^{-2/3} ds^2 + ds_K^2, \label{uplift}
\ee
where the warp factor $\Delta$ depends on the sphere coordinates and $ds_K^2$ is a metric on a warped sphere. Explicitly
\bea
\Delta^{-2/3} &=& \frac{\zeta}{\lambda^2} \qquad \zeta = (1 + \frac{\sigma^2}{\rho^2} \cos^2 \theta); \\
ds_K^2 &=& \frac{1}{\zeta} \left ( \zeta^2 d \theta^2 + \cos^2 \theta d\Omega_3^2 + \lambda^6 \sin^2 \theta d \phi^2 \right ). \nn
\eea
The five-dimensional metric has a naked timelike singularity at $\rho = 0$ but the uplifted geometry has a null horizon at $\rho =0$. 

\begin{figure}
\begin{center}
\setlength{\unitlength}{0.80mm}
\includegraphics*[trim =2.8cm 11.2cm 0.5cm 12cm,width=1.2\linewidth]{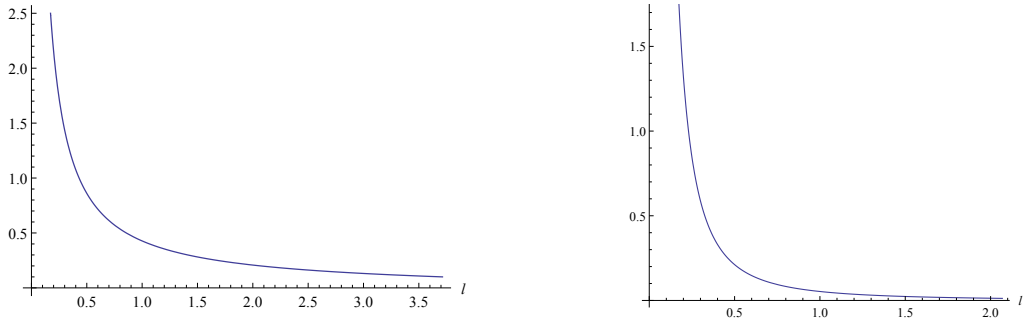}
\caption{Plots of $\rho_0$ and $\Delta S$ as functions of $l$ for $\sigma =0.1$. Both functions asymptote to zero as $l \rightarrow \infty$.}
\label{deltaf}
\end{center}
\end{figure} 

The five-dimensional metric satisfies an a-theorem: the warp factor $A(r)$  in \eqref{5dgeo} decreases monotonically as $r$ decreases. Correspondingly the entanglement entropy and the differential entropy monotonically decrease as the scale of the entangling region is increased. Working for convenience in the coordinate system \eqref{ocs} the entanglement entropy of a slab is 
\be
S = \frac{L^2}{4 G_N} \int_{\rho_0}^{\Lambda} d\rho \frac{\rho^3 (\rho^2 + \sigma^2)^{1/2}}{\sqrt{\rho^4 (\rho^2 + \sigma^2) - \rho_0^4 (\rho_0^2 + \sigma^2)}},
\ee
where $\Lambda$ is the UV cutoff and $\rho_0$ is the turning point. The quantity 
\be
\Delta S = S - \frac{\Lambda^2 L^2}{2 G_N}
\ee
is UV finite by construction. The relation between the width of the slab $l$ and the turning point is 
\be
l = 2 \rho_0^2 (\rho_0^2 + \sigma)^{1/2} \int^{\infty}_{\rho_0} \frac{d \rho}{\sqrt{\rho^2(\rho^2 + \sigma^2)(\rho^4 (\rho^2 + \sigma^2) - \rho_0^4 (\rho_0^2 + \sigma^2))}}. 
\ee
The quantity $\Delta S(l)$ monotonically decreases as $l$ is increased. We can express the differential entropy associated with strips of width $l$ as 
\be
E (l) = L_x \frac{\partial (\Delta S)}{\partial l},
\ee
and this quantity also decreases monotonically with $l$. This had to be true since (by construction) $E(l)$ can be expressed in terms of the warp factor in \eqref{5dgeo}, which is known to satisfy the a-theorem, see \cite{Freedman:1999gp}. It is interesting to note that the differential entropy is proportional to the finite entropy \eqref{sl}, which in turn is known to play the role of a c-function in two spacetime dimensions. In any holographic background dual to a RG flow, the differential entropy is by construction expressed in terms of the warp factor in \eqref{5dgeo}, which satisfies the a-theorem provided that appropriate energy conditions are imposed on the bulk stress energy tensor \cite{Freedman:1999gp}, and therefore the differential entropy has the correct property to correspond to an a-function (in any dimension). 

\bigskip

Entanglement for Coulomb branch geometries has been discussed in earlier papers \cite{Mollabashi:2014qfa,Aprile:2014iaa} from a ten-dimensional perspective, i.e. minimal surfaces in the ten dimensional geometry were explored. As emphasised throughout this paper, the standard entanglement entropy can only be computed from the five-dimensional Einstein metric, which can only be extracted near the boundary in the case of separated brane stacks discussed in  \cite{Mollabashi:2014qfa,Aprile:2014iaa}, since no consistent truncation to five-dimensional supergravity exists (Kaluza-Klein holography allows us to extract the Einstein metric near the conformal boundary, where the metric is close to $AdS_5$ but deep in the interior the metric is not close to $AdS_5$ and therefore the method cannot be used there, see \cite{Skenderis:2006uy,Skenderis:2006di}). Codimension two minimal surfaces in the uplifted geometry 
\eqref{uplift}  compute the generalized entanglement entropy defined in  \cite{Mollabashi:2014qfa}. It would be interesting to understanding the field theory interpretation of this quantity better; see recent discussions in \cite{Karch:2014pma}. 

\section{Conclusions} \label{nine}

One of the main results of this paper is a systematic method to compute the effective lower-dimensional Einstein metric for top-down brane probe systems, using which one can extract the entanglement entropy. We have illustrated this method with the case of quarks at finite mass and density, at zero temperature. It would be interesting to explore the finite temperature behaviour of the entanglement entropy and how it captures the phase transitions found in \cite{Mateos:2006nu,Kobayashi:2006sb}. Our method is applicable to any brane probe system and could for example be used to evaluate the entanglement entropy in models of quantum Hall physics \cite{Bergman:2010gm,Kristjansen:2012ny} and in top-down models of the Kondo effect \cite{Erdmenger:2013dpa}. One could also explore entanglement entropy in the presence of flavors for ABJM theory, see earlier results in  \cite{Bea:2013jxa,Kim:2014yca}.

The entanglement entropy calculated using the ten-dimensional metric is in general qualitatively different to that computed using the lower-dimensional Einstein metric. It is interesting to note however that the entanglement entropy computed using the top down metric for smeared solutions such as \cite{Bea:2013jxa} seems to give answers which agree with F-theorem expectations. In such examples the ten-dimensional metric has a very special form, in which all warp factors depend only on the radial coordinate, and thus the lower-dimensional Einstein metric is simply related to the top down metric. It would be interesting to understand the relationship between the metrics in more detail, and to compare the entanglement entropy computed in \cite{Bea:2013jxa} with what is obtained using the method developed here. 

The entanglement entropy is not the only interesting quantity which is computable from the effective lower-dimensional Einstein metric: correlation functions involving only the transverse traceless components of the field theory stress energy tensor can also be computed from perturbations of the Einstein metric. Kaluza-Klein holography allows such energy momentum tensor correlation functions to be accessed without computing the entire backreaction in ten dimensions. 

We were able to match the structure of all terms in the entanglement entropy for finite mass quarks at zero density with field theory expectations, and the logarithmic divergences were matched exactly. Few field theory results exist for entanglement entropy at finite temperature and density; see the recent papers 
\cite{Cardy:2014jwa,Herzog:2014fra} for discussions of specific universal thermal corrections in conformal field theories. It would be interesting to explore whether conformal perturbation techniques analogous to those of \cite{Rosenhaus:2014zza} can be used to extract universal terms at finite mass and density; just like the mass, the chemical potential breaks conformal invariance even at zero temperature but can be treated in conformal perturbation theory. Results on entanglement entropy using conformal perturbation theory in the context of higher spin theory can be found in \cite{Datta:2014ska,Datta:2014uxa,Datta:2014zpa}. 

In this work we have emphasised that the entanglement entropy and differential entropy are associated with the lower-dimensional Einstein metric, rather than the ten-dimensional metric, and that the latter can only be reconstructed given additional information. The emergence and reconstruction of the compact part of the bulk geometry is a longstanding puzzle. If entanglement can indeed be used to reconstruct the bulk spacetime, then there must exist in the field theory a generalised measure of entanglement which captures the compact part of the geometry. Attempts to define such a quantity were made in \cite{Mollabashi:2014qfa,Karch:2014pma} and we will report elsewhere on extensions of these proposals.

\section*{Acknowledgments}

We would like to thank Kostas Skenderis for collaboration in early stages of this work. 
MMT acknowledges support from a grant of the John Templeton Foundation. The opinions expressed in this publication are those of the authors and do not necessarily reflect the views of the John Templeton Foundation. 
This work was supported by the Science and Technology Facilities Council (Consolidated Grant ``Exploring the Limits of the Standard Model and Beyond'') and by the Engineering and Physical Sciences Research Council. 
This work was supported in part by National Science Foundation Grant No. PHYS-1066293 and the hospitality of the Aspen Center for Physics.
We thank the Galileo Galilei Institute for Theoretical Physics for the hospitality and the INFN for partial support during the completion of this work.

\appendix

\section{Source terms in linear equations} \label{ten}

In this appendix we discuss the derivation of \eqref{src1} from \eqref{iib-corr} using the results for the linearised field equations around an $AdS_5 \times S^5$ background given in 
\cite{Kim:1985ez}.  The components of the Einstein equations in the non-compact directions are
\bea
&& \frac{1}{2}  ( \Box_x + \Box_y + 2 ) H_{mn}  + 3 g^o_{mn} H_{p}^p - \frac{1}{2} \nabla_m \nabla^p H_{np} - \frac{1}{2} \nabla_n \nabla^p H_{mp} \\
&& \qquad 
+ \frac{1}{2} \nabla_{m} \nabla_n H^p_p - \frac{1}{2} \nabla_m \nabla^a h_{n a} - \frac{1}{2} \nabla_n \nabla^a h_{ma} - \frac{1}{6} g^o_{mn} (\Box_x + \Box_y) h^a_{a} \nn \\
&& \qquad
- \frac{16}{3} g^o_{mn} h^{a}_a + \frac{1}{3} g^{o}_{mn} \epsilon^{pqrst} f_{pqrst} = \kappa_{10}^2 \bar{T}_{mn}. \nn
\eea
Here $\nabla_m$ and $\nabla_a$ denote covariant derivatives while $\Box_x$ and $\Box_y$ denote the d'Alambertians; $H_{np} = h_{np} + \frac{1}{3} h^{a}_{a} g^o_{mn}$.  

Projecting this equation onto the zeroth spherical harmonic results in 
\bea
&& \frac{1}{2} (\Box_x + 2) h^E_{mn} + 3 g^o_{mn} (h^E)^p_p  - \frac{1}{2} \nabla_m \nabla^p h^E_{np} - \frac{1}{2} \nabla_n \nabla^p h^E_{mp}  \label{el0} \\
&& \qquad
+ \frac{1}{2} \nabla_{m} \nabla_n (h^E)^p_p - \frac{1}{6} g^o_{mn} \Box_x \pi^0 
- \frac{16}{3} g^o_{mn} \pi^0  + \frac{1}{3} g^{o}_{mn} \epsilon^{pqrst} \partial_p b^0_{pqrst} = \kappa_{10}^2 \bar{T}^0_{mn}, \nn
\eea
where $h^{E}_{mn}$ was defined in \eqref{ein-def}. 

To obtain an equation for the Einstein metric perturbation $h^{E}_{mn}$ we need to eliminate $\pi^0$ and $b^{0}_{pqrs}$. The trace of the Einstein equation over the five sphere gives
\bea
&& \frac{1}{2} (\Box_x - \frac{1}{15} \Box_y  - 32) h^{a}_a  + \frac{1}{2} \Box_y H^{p}_p - \frac{1}{2} \nabla^a \nabla^p h_{ap} \\
&& \qquad
- \nabla^a  \nabla^b h_{(ab)} + \frac{5}{3} \epsilon^{abcde} f_{abcde} = \kappa_{10}^2 \bar{T}_{a}^a. \nn
\eea
Projecting this equation onto the zeroth spherical harmonic results in 
\be
\frac{1}{2} (\Box_x - 32) \pi^0 = \kappa_{10}^2 \tilde{T}^0.  \label{el1}
\ee
The five-form self duality equation along the non-compact directions gives
\be
5 \partial_{ [ m} c_{npqr] } = \frac{1}{4!} \epsilon_{mnpqr}^{\; \; \; \; \; abcde} \partial_a c_{bcde} + \frac{1}{2} (H^{p}_p - \frac{8}{3} h^a_a) \epsilon_{mnpqr}
\ee
which projected onto the zeroth spherical harmonics gives
\be
5 \partial_{ [ m} b^0_{npqr]} =  \frac{1}{2} ((h^E)^p_p - \frac{8}{3} \pi^0) \epsilon_{mnpqr} \label{el2}
\ee
Inserting \eqref{el1} and \eqref{el2} into \eqref{el0} then gives
\be
( {\cal L}_E + 4 ) h^E_{mn} = \kappa_{10}^2 ( \bar{T}^0_{mn} + \frac{1}{3} \tilde{T}^0 g^o_{mn})
\ee




\providecommand{\href}[2]{#2}\begingroup\raggedright\endgroup


\end{document}